\documentclass[epj]{svjour}
\usepackage{graphics}
\usepackage{inputenc}
\usepackage{epsfig,float}
\usepackage{amsfonts}
\usepackage{amsmath}
\newcommand{\im}{{\rm Im}}
\newcommand{\re}{{\rm Re}}

\newcommand{\be}{\begin{eqnarray}}
\newcommand{\ee}{\end{eqnarray}}

\begin{document}
\title{Dual parametrization of GPDs versus the double distribution Ansatz}
\author{Maxim V. Polyakov\inst{1,2}
\thanks{Maxim.Polyakov@tp2.ruhr-uni-bochum.de}
\and Kirill M. Semenov-Tian-Shansky\inst{1,3}
\thanks{semenovk@tp2.ruhr-uni-bochum.de}%
}                     
\offprints{}          
\institute{Institut f\"{u}r Theoretische Physik II, Ruhr-Universit\"{a}t Bochum, D-44780 Bochum, Germany
\and
Petersburg Nuclear Physics Institute, Gatchina, St. Petersburg 188350, Russia
\and
St. Petersburg State University, St.Petersburg, Petrodvoretz, 198504,  Russia}
\date{Received: date / Revised version: date}
%
\abstract{
We establish a link between the dual parametrization of GPDs
and a popular parametrization based on the double distribution Ansatz,
which is in prevalent use in phenomenological applications.
We compute several first forward-like functions that 
express the
double distribution Ansatz for GPDs in the framework of the dual
parametrization and show that these forward-like functions make
the dominant contribution into the GPD quintessence function.
We also argue that the forward-like functions
$Q_{2 \nu}(x)$
with
$\nu \ge 1$
contribute to the leading singular small-$x_{Bj}$
behavior of the imaginary part of DVCS amplitude.
This makes the small-$x_{Bj}$
behavior of
$\im A^{DVCS}$
independent of the asymptotic behavior of PDFs. Assuming analyticity of
Mellin moments of GPDs in the Mellin space
we are able to fix the value of the $D$-form factor in terms of the GPD
quintessence function $N(x,t)$ and the forward-like function $Q_0(x,t)$.
} 

\authorrunning{M. Polyakov and K. Semenov-Tian-Shansky }
\titlerunning{Dual parametrization of GPDs {\it vs.} the double distribution Ansatz}

\maketitle

\section{Introduction}

Generalized parton distributions (GPDs)
\cite{pioneers}
have been investigated in great details during the past decade.
These distributions, proved to be extremely efficient for the
description of the quark and gluon structure of hadrons.
GPDs provide a natural generalization of parton distribution functions,
familiar from inclusive reactions, and elastic form-factors.
The current understanding of both theoretical and experimental aspects of GPDs
is reviewed in refs.
\cite{GPV,Diehl,BelRad,Boffi:2007yc}.
Extraction of GPDs from the experimental data is highly
demanded, since these functions contain valuable new information on hadron
structure.
The theoretical opportunity to access GPDs experimentally is provided by
the collinear factorization theorems for hard exclusive reactions
\cite{Collins:1998be,Collins:1996fb}.
The dedicated experiments
\cite{DVCSexp}
performed during the last years provide an increasing amount of precise
experimental data.  Unfortunately, since GPDs  are complicated functions of the
longitudinal momentum fraction of partons
($x$),
skewness parameter
($\xi$),
the momentum transfer squared
($t$)
as well as of factorization scale, the direct extraction of GPDs from
the observables turns out to be an extremely difficult task. Moreover,
GPDs always enter the observable quantities being integrated over
$x$
with weighting functions given by the propagators of partons between
the incoming virtual photon and the outgoing real photon or meson.
Therefore, in order to extract GPDs from the data, one usually rely on
different phenomenologically motivated parameterizations and simultaneous
fitting procedures for several observables.

An important advantage of the dual parametrization of GPDs
\cite{Polyakov:2002wz}
is that it suggests the possible form of deconvolution procedure
\cite{Tomography,Polyakov:2007rw,Moiseeva:2008qd}
which allows to specify the maximum amount of information on GPDs that can
be extracted from the experimental data in a single GPD quintessence function.
GPD quintessence function can be unambiguously recovered from the leading
order amplitude with the help of the Abel transform tomography method. Another
gain from the dual parametrization is that it allows one to distinguish
the contribution to the observables brought by parton densities from the
genuine non-forward contributions (see discussion in
\cite{Polyakov:2008xm}).

In this paper, using reparametrization procedure described in
\cite{ForwardLikeF_KS},
we establish the link between the dual parametrization of GPDs and
the famous Radyushkin double distribution Ansatz
\cite{Radyushkin:1997ki,Radyushkin:1998bz,RadDDandEvolution,Musatov:1999xp,Radyushkin:2000uy}
employed in numerous phenomenological applications. The explicit expressions
for the first forward-like functions
$Q_2$
and
$Q_4$
allow us to quantify the relative importance of non-forward effects encoded
in these functions. We also argue that, in the framework of the dual
parametrization, the forward-like functions
$Q_{2 \nu}(x)$
with
$\nu \ge 1 $
may contribute to the leading singular small-$\xi$
asymptotic behavior of the imaginary part of the leading order DVCS amplitude.
In this case the forward-like functions
$Q_{2 \nu}(x)$
with
$\nu \ge 1 $
may have the small-$x$
singularities, which lead to divergencies of the generalized form-factors
$B_{2\nu-1 \,\; 0}$.
We discuss the regularization procedure that allows to assign a precise meaning
to the potentially divergent integrals for generalized form factors
$B_{2\nu-1 \,\; 0}$.
Assuming the analyticity of Mellin moments of GPD in Mellin space
(see \cite{Mueller:2005ed,Kumericki:2007sa,Kumericki:2008di})
one can unambiguously fix the contribution of the
$D$-form factor into the real part of the DVCS amplitude in terms of the
GPD quintessence function
and the forward-like function
$Q_0$.

Finally, in the Appendix~\ref{App_A}
in order to avoid
confusions in the literature
we review the form of the integral transformation
\cite{Polyakov:2002wz,Tomography}
relating GPD and the set of forward-like functions
$Q_{2 \nu}$
in the framework of the dual parametrization. We generalize this result for
the case when  the forward-like functions
$Q_{2 \nu}(x)$
with
$\nu \ge 1 $
are allowed to have
small-$x$
singularities, which lead to divergencies of the generalized form-factors
$B_{2\nu-1 \,\; 0}$.

\section{Basic facts on the dual parametrization of GPDs}

The dual parametrization of GPDs
\cite{Polyakov:2002wz}
is based on the representation of
GPDs as infinite sums of $t$-channel resonance exchanges
\cite{Polyakov:1998ze}.
Originally, the dual parametrization was formulated for the case of spinless
hadrons. First we briefly  discuss the difference, which arises when
dealing with spin one half particles (a detailed discussion on this issue
is presented in \cite{Moiseevaetal}).
Note that the specifics related to spin-$\frac{1}{2}$ was not taken into account in the
early phenomenological applications of the dual parametrization
\cite{Guzey:2006xi}, as it has been stressed in \cite{Diehl:2008tn}.

According to
\cite{Diehl,Diehl:2007jb},
the following (so-called
electric and magnetic)
combinations of nucleon GPDs
$H^q(x,\xi,t)$
and
$E^q(x,\xi,t)$
are suitable for partial wave expansion in the $t$-channel:
\begin{equation}
\begin{split}
& H^{(E)}(x,\xi,t)=H^q(x,\xi,t)+\frac{t}{4m_N^2} E^q(x,\xi,t)\, , \\&
  H^{(M)}(x,\xi,t)=H^q(x,\xi,t)+E^q(x,\xi,t)\,.
\end{split}
\end{equation}
Here we employ the nucleon generalized parton distribution
of the particular flavor
$q$:
$H^q(x,\xi,t)$,
$E^q(x,\xi,t)$
with the properties listed below.
\begin{itemize}
\item Generalized parton distributions
$H^q(x,\xi,t)$,
$E^q(x,\xi,t)$
defined for
$x \in [-1,1]$
are reduced to the following $t$-dependent
quark densities in the limit $\xi \rightarrow 0$:
\be
H^q(x, \xi=0,t)=
\begin{cases}
q(x,t) \ \ \ \text{for} \ \ \ x>0 \\
-\bar{q}(-x,t) \ \ \ \text{for} \ \ \ x<0
\end{cases}\,;
\ee
\be
E^q(x, \xi=0,t)=
\begin{cases}
e^q(x,t) \ \ \ \text{for} \ \ \ x>0 \\
-\bar{e}^q(-x,t) \ \ \ \text{for} \ \ \ x<0
\end{cases}\,;
\ee

\item The following normalization conditions are fulfilled:
\be
&& \int_{-1}^1 dx \,x H^{(E)}(x, \xi,t=0)
= M_2^q+\frac{4}{5} d_1^q \, \xi^2\,; \nonumber \\ &&
\int_{-1}^1 dx \,x H^{(M)}(x, \xi,t=0)=2J^q\,,
\ee
where
$M_2^q$
stands for the momentum fraction carried by quarks and antiquarks of
the particular flavor $q$ in the nucleon;
$J^q$ denotes angular momentum carried by quarks of flavor $q$;
$d_1^q$ is the first coefficient of the Gegenbauer expansion of the $D$-term
of flavor $q$.
\end{itemize}

The dual parametrization can be introduced for the
singlet and nonsinglet combinations of
$H^{(E,M)}$.
Below we consider only the case of the singlet ($C=+1$)
combinations of
$H^{(E,M)}$,
which enter the description of DVCS.
These combinations are introduced according to
\be
&& H^{(E)}_+(x,\xi,t)=H^{(E)}(x,\xi,t)-H^{(E)}(-x,\xi,t)\,;
\nonumber
\\&&
H^{(M)}_+(x,\xi,t)=H^{(M)}(x,\xi,t)-H^{(M)}(-x,\xi,t)\,.
\ee
Note that the singlet electric and magnetic combinations of
nucleon GPDs
$H^{(E,M)}(x,\xi,t)$
are defined for
$x \in [0,\,1]$.
For
$\xi=0$
they are reduced to the following combinations of
$t$-dependent quark densities
$q_+(x,t) \equiv q(x,t)+\bar{q}(x,t)$,
$e_+^q(x,t)  \equiv e^q(x,t)+\bar{e}^q(x,t)$:
\be
&& H^{(E)}_+(x,\xi=0,t)=q_+(x,t)+ \frac{t}{4 m_N^2}\, e_+^q(x,t)\,;  \label{Limit_xi_0_singlet}  \\&&
H^{(M)}_+(x,\xi=0,t)=q_+(x,t)+  e_+^q(x,t)\,.
\label{Limit_xi_0_singlet2}
\ee
The singlet electric and magnetic combinations of
nucleon GPDs
$H^{(E,M)}_+(x,\xi,t)$
are normalized according to
\be
&& \int_{0}^1 dx \,x H^{(E)}_+(x, \xi,t=0)= M_2^q+\frac{4}{5} d_1^q \, \xi^2\,; \nonumber \\ &&
\int_{0}^1 dx \,x H^{(M)}_+(x, \xi,t=0)=2J^q\,.
\ee

All formulae for the electric combination  of nucleon GPDs are the same as in the
case of the scalar hadrons addressed in
\cite{Polyakov:2002wz}.
Thus, the partial wave decomposition in the $t$-channel for the singlet
electric combination of GPDs
$H^{(E)}_+(x, \xi, t)$
is written as the following formal series%
\footnote{Note that we have introduced an additional factor $2$ in the
partial wave expansions
(\ref{FSGPD}),
(\ref{Formal_sum_H_M_odd_in_x}).
This is done to make the resulting
GPDs $H^{(E,M)}_+$ satisfy
(\ref{Limit_xi_0_singlet}),
(\ref{Limit_xi_0_singlet2})
in the limit
$\xi \rightarrow 0$. In the original paper
\cite{Polyakov:2002wz}
a rather unusual convention that differs
by a factor
$\frac{1}{2}$
from
(\ref{Limit_xi_0_singlet})
was employed. In particular this unusual convention has led to
much confusion in the early phenomenological applications of
the dual parametrization
\cite{Guzey:2005ec,Guzey:2006xi}
(see  \cite{Guzey:2008ys} for the detailed discussion of this issue).
}:
\be
&& H^{(E)}_+(x, \xi, t)=
\nonumber
\\ &&
2 \sum_{n=1 \atop \text{odd}}^\infty
\sum_{l=0 \atop \text{even}}^{n+1}
B_{nl}^{(E)}(t) \,
\theta
\left(
1-\frac{x^2}{\xi^2}
\right)
\left(
1-\frac{x^2}{\xi^2}
\right)
C_n^{\frac{3}{2}}
\left( \frac{x}{\xi} \right)
P_l
\left( \frac{1}{\xi} \right), \nonumber
\\ &
\label{FSGPD}
\ee
where
$C_n^{\frac{3}{2}}(\chi)$
stand for the Gegenbauer polynomials;
$P_l(\chi)$
are Legendre polynomials; 
$B_{nl}(t)$
are the generalized form factors;
$x$, $\xi$
and
$t$
stand for usual GPD variables.
As it is pointed out in
\cite{Moiseevaetal},
the $t$-channel
partial wave expansion for the magnetic combination
goes over
$\frac{1}{\xi} P'_l \left( \frac{1}{\xi} \right)$:
\be
&&
H^{(M)}_+(x, \xi, t)=
\nonumber \\&&
2 \sum_{n=1 \atop {\rm odd}}^\infty
\sum_{l=0 \atop {\rm even}}^{n+1}
B_{nl}^{(M)}(t) \,
\theta
\left(
1-\frac{x^2}{\xi^2}
\right)
\left(
1-\frac{x^2}{\xi^2}
\right)
C_n^{\frac{3}{2}}
\left( \frac{x}{\xi} \right)
\frac{1}{\xi }
P'_l
\left( \frac{1}{\xi } \right).
\nonumber \\ &&
\label{Formal_sum_H_M_odd_in_x}
\ee

In the following we consider only the case of the singlet
($C=+1$)
electric nucleon
GPD. The variable $t$
plays no particular role for our analysis, so for simplicity we
set it to zero for the rest of this section.
 In this limit the singlet electric combination of nucleon GPDs
is reduced to the usual $C=+1$ nucleon GPD $H_+$.
Thus,
we omit the superscript ``$(E)$'':
$$
H^{(E)}_+(x,\xi,t=0) \equiv H_+(x,\xi) \,.
$$
In the forward limit $H_+(x, \xi)$ is reduced to
$q_+(x)$,
where
$q_+(x)=q(x)+\bar{q}(x)$
stands for the singlet combination of forward quark distributions. However,
the generalization of our analysis (in particular, of the main results of
sects.~\ref{Sec_toy}, \ref{Sec_Q2Q4_for_double})
for fixed
$t \ne 0$
is straightforward.

Below, we list a number of important facts concerning the dual
parametrization of GPDs.
\begin{itemize}
\item
In order to sum up the formal series
(\ref{FSGPD}) for
$H_+(x, \xi)$
(see  Appendix~\ref{App_A})
one has to introduce
the set of {\em forward-like functions}
$Q_{2 \nu}(x)$ $(\nu=0, \,1,\,...)$,
whose Mellin moments generate the generalized form factors
$B_{nl}$:
\begin{equation}
 B_{n \, n+1-2 \nu}= \int_0^1 dx x^n Q_{2 \nu}(x)\,.
 \label{Bnl_int}
\end{equation}
\item At the LO the scale dependence of the forward-like functions
$Q_{2 \nu}(x)$
is given by the standard DGLAP evolution equation so that these functions
evolve as usual parton distributions under QCD evolution.
\item The forward-like function
$Q_0(x)$
is related to the forward quark densities according to
\begin{equation}
Q_0(x)=q_+(x)- \frac{x}{2} \int_x^1   \frac{dy}{y^2} \,  q_+(y)\, .
\label{Q0final}
\end{equation}
\item The leading order twist-$2$ amplitude
$A(\xi) \equiv A(\xi,t=0)$:
\begin{equation}
A(\xi)= \int_0^1 dx \, H_+(x,\xi)
\left[
\frac{1}{\xi-x-i \epsilon}- \frac{1}{\xi+x-i \epsilon}
\right]
\label{DVCS_ampl}
\end{equation}
is completely determined
by the
GPD-quintessence function
$N(x)$
\cite{Tomography,Polyakov:2007rw}:
\begin{equation}
N(x)=\sum_{\nu=0}^\infty x^{2 \nu} Q_{2 \nu}(x)
\label{Quintes}
\end{equation}
and by the $D$-form factor
$D^q$,
that is given by
\begin{equation}
D^q= \frac{1}{2} \int_{-1}^1 dz \, \frac{D^q(z)}{1-z}= \sum_{n=1 \atop \text{odd}}^\infty d_n^q \,,
\label{Dff}
\end{equation}
where
\begin{equation}
D^q(z)=
(1-z^2) \sum_{n=1 \atop \text{odd}}^\infty d_n^q C_n^{3/2}(z)
\label{Geg_exp_D}
\end{equation}
stands for the $D$-term.
The calculations performed in the chiral quark-soliton
model
\cite{Petrov:1998kf}
suggest
$D^u(z)\approx D^d(z)$,
thus
$D^q(z)=\frac{1}{N_f} D(z)$.
$N_f$
stands here for the number of
flavors
and
$D(z)= (1-z^2) \sum_{n, \, \text{odd}} d_n C_n^{3/2}(z)$
is the flavor singlet $D$-term.

\item The partial wave expansion
of
the DVCS amplitude
(\ref{DVCS_ampl})
reads
\cite{Polyakov:2002wz}
\be
A(\xi)= 4 \sum_{n=1 \atop \text{odd}}^\infty
\sum_{l=0 \atop \text{even}}^{n+1} B_{n\;l}
P_l \left( \frac{1}{\xi} \right)\,.
\ee
This formal series can be summed up yielding the following
result for the DVCS amplitude:
\be
&& A(\xi)=2
\int_0^1 \frac{dx}{x} \sum_{\nu=0}^\infty
N(x)  \nonumber \\ &&
\times
\left[
\frac{1}{\sqrt{1-\frac{2x}{\xi}+x^2}}
+
\frac{1}{\sqrt{1+\frac{2x}{\xi}+x^2}}
-2
\right]
\nonumber \\ &&
+ 4 \, \int_0^1 \frac{dx}{x} (N(x)-Q_0(x))\,.
\label{DVCSampl_added}
\ee

\item The explicit expressions for the imaginary and real parts of
the DVCS amplitude
(\ref{DVCS_ampl})
read
\cite{Polyakov:2002wz,Tomography}
\begin{equation}
\text{Im} A(\xi)= 
2
\int_{\frac{1- \sqrt{1-\xi^2}}{\xi}}^1
\frac{dx}{x} \, N(x)
\left[
\frac{1}{\sqrt{\frac{2x}{\xi}-x^2-1}}
\right]\,; 
\label{ImA_Dual}
\end{equation}
\begin{equation}
\begin{split}
& \text{Re} A(\xi) =  4 D^q+
 2 \int_0^{\frac{1- \sqrt{1-\xi^2}}{\xi}}
\frac{dx}{x} \, N(x) 
\left[
\frac{1}{\sqrt{1- \frac{2x}{\xi}+x^2}} \right. \\&
\left.
+
\frac{1}{\sqrt{1+ \frac{2x}{\xi}+x^2}}-
\frac{2}{\sqrt{1 +x^2}}
\right] \\&
+2 \int_{\frac{1- \sqrt{1-\xi^2}}{\xi}}^1
\frac{dx}{x} \, N(x)
\left[
\frac{1}{\sqrt{1+ \frac{2x}{\xi}+x^2}}
-
\frac{2}{\sqrt{1 +x^2}}
\right]\,.
  \end{split}
  \label{ReA_Dual}
\end{equation}

\item The $D$-form factor $D^q$ can be formally%
\footnote{The problem of treating the possible divergencies in
(\ref{D_form_factor_NQ}) is discussed in sect.~\ref{Sec_toy}.}
expressed through the GPD quintessence function
$N(x)$
and the forward-like function
$Q_0$
according to
\cite{Tomography}
\begin{equation}
\begin{split}
D^q=&
\int_0^1 \frac{dx}{x}N(x)
\left(
\frac{1}{\sqrt{1+x^2}}
-1
\right) \\&
+\int_0^1 \frac{dx}{x}
\left(
N(x)-Q_0(x)
\right)
\,.
\end{split}
\label{D_form_factor_NQ}
\end{equation}

\item
The polynomiality condition for Mellin moments of generalized parton distribution
\cite{pioneers}
implies that for
$N=1,\,3,\,... \,$:
\begin{equation}
\int_{0}^1 dx \, x^N H_+(x,\xi)= 
  h_0^{(N)}+h_2^{(N)} \xi^2+...+h_{N+1}^{(N)} \xi^{N+1}  \, .
\label{Mellin_M}
\end{equation}
In the framework of the dual parametrization,
the set of coefficients
$h_{2 \nu}^{(N)}$
with
$ \nu \le \frac{N+1}{2}$
is given by
\be
  &&
  h_{2 \nu}^{(N)}=\sum_{n=1 \atop \text{odd}}^N
  \sum_{l=0 \atop \text{even}}^{n+1}
  B_{nl} \,
  {(-1) }^{\frac{2 \nu +l-N-1 }{2}}  \nonumber \\ &&
  \times
    \frac{\Gamma (1 - \frac{2 \nu - l - N}{2})}{\Gamma (\frac{1}{2} + \frac{2 \nu + l - N}{2})\,\Gamma (2 - 2 \nu + N)\,}
   \nonumber  \\ &&
       \times
    \frac{\left(  n+1 \right) \,\left(  n+2 \right) \,\Gamma ( N+1)\,
    }{2^{2 \nu}\,
    \Gamma (1 + \frac{N-n}{2})\,\Gamma (\frac{5}{2} + \frac{N + n}{2})}\,.
%
\label{hk_dual}
\ee
 Note that due to the Gamma function in the denominator only the
terms with $l \ge N+1-2 \nu$ are non zero in the sum over $l$ in
(\ref{hk_dual}).
As the result for the given odd
$N$
and
$\nu$
($\nu \le \frac{N+1}{2}$)
only the forward-like
functions
$Q_{2 \mu}$
with
$\mu=0,\,1,\,...\,, \nu$ are relevant
for the computation of the coefficient
$h_{2 \nu}^{(N)}$.

\item An important role is played by the expansion of GPD
$H_+(x, \xi)$
in powers of
$\xi$
around the point
$\xi=0$
with fixed
$x$ ($x>\xi$).
To the order
$\xi^2$
this expansion is given by
\cite{Tomography}
\begin{equation}
\begin{split}
& H_+(x,\xi)= H_+^{(0)}(x)+ \xi^2 H_+^{(1)}(x)+ \xi^4 H_+^{(2)}(x)+... \\&
=Q_0(x)+
\frac{\sqrt{x}}{2}
\int_x^1 \,
\frac{dy}{y^{3/2}} \, Q_0(y) \\&
+\xi^2
\left[
-\frac{1-x^2}{4x}
\frac{\partial }{\partial x}Q_0(x)+
\right.
\\&
\left.
\frac{1}{32}
\int_x^1 dy \, Q_0(y)
\left\{
\frac{1}{y} \left(
3\,{\sqrt{\frac{x}{y}}} + 3\,{\sqrt{\frac{y}{x}}}
\right)
\right.
\right.
\\&
\left.
\left.
+
\frac{1}{y^3} \left(
3\,{\sqrt{\frac{y}{x}}} - {\left( \frac{y}{x} \right) }^{\frac{3}{2}}
\right)
\right\}+
\frac{1}{4}
Q_2(x)
\right. \\&
\left.
+
%
\frac{3}{32}
\int_x^1
dy \, Q_2(y)
\frac{1}{y}
\left(
\frac{1}{2}{\sqrt{\frac{x}{y}}} + {\sqrt{\frac{y}{x}}} +
  \frac{5}{2} {\left( \frac{y}{x} \right) }^{\frac{3}{2}}
\right)
\right]+ O(\xi^4)
\end{split}
\label{small_xi_dual}
\end{equation}
\end{itemize}
The result up to the order
$\xi^4$
was derived in
\cite{ForwardLikeF_KS}.
A remarkable property of such expansion is that up to the
particular order
$\xi^{2 \nu}$
it involves only a finite number of forward-like functions
$Q_{2 \mu}(x)$
with
$\mu \le \nu$.
This property allows to invert the expansion
(\ref{small_xi_dual})
and to express the set of
forward-like functions through GPDs for various phenomenological
parametrizations of GPDs.
This enables us to rewrite the particular phenomenological parametrizations of GPDs in
the framework of the dual parametrization
\cite{ForwardLikeF_KS}.
Let us assume that the expansion of GPD
$H_+(x, \xi)$
around
$\xi= 0$
for
$x>\xi$
calculated in the framework of a certain
model is known:
\begin{equation}
H_+(x, \xi)=
\phi_0(x)+\phi_2(x) \xi^2 + \phi_4(x) \xi^4+O(\xi^6)\,,
\label{Expansion_start}
\end{equation}
where
$$
\phi_{2 \nu}(x)= \frac{1}{(2\nu)!}\,
\frac{\partial^{2 \nu}  } {\partial \xi^{2\nu}}\,
H_+(x, \xi)_{\xi=0}\,.
$$
Below we list the explicit expressions for several
first forward-like functions
$Q_2(x)$
and
$Q_4(x)$
derived in \cite{ForwardLikeF_KS}
from matching the expansions
(\ref{small_xi_dual})
and
(\ref{Expansion_start}).
Clearly, since the GPD calculated in the realistic model is supposed to
have the correct forward limit
$$
\phi_0(x) \equiv H_+(x, \xi=0)=q_+(x)\,, 
$$
the usual result
(\ref{Q0final})
for
$Q_0(x)$
is recovered.
For
$Q_2(x)$
such expression reads
\be
&& Q_2(x)=
\frac{2(1-x^2)}{x^2} \, q_+(x)+
\frac{(1-x^2)}{x} \, q'_+(x) \nonumber \\ && +
\int_x^1 dy \left(
\frac{-15\,x}{4\,y^4} - \frac{3}{2\,y^3} +
  \frac{5\,x}{4\,y^2}\right)q_+(y) \nonumber \\ &&
+ 
4\, \phi_2(x)-
  \int_x^1 dy \,
  \phi_2(y)
\left(
  \frac{15\,x}{4\,y^2} + \frac{3}{2y}+\frac{3}{4\,x}
\right).
\label{Q2final}
\ee
The explicit expression
for
$Q_4(x)$
reads
\begin{equation}
\begin{split}
& Q_4(x) =
\left( \frac{49}{8\,x^4} - \frac{17}{8\,x^2} \right) \,\left( 1 - x^2
\right)q_+(x) \\& +
\left( \frac{7}{4\,x^3} - \frac{11}{4\,x} \right) \,\left( 1 - x^2 \right)
q_+'(x)+ \frac{{\left( 1 - x^2 \right) }^2}{2\,x^2} q_+''(x)
\\&
-\int_x^1dy \, q_+(y)
\left(
\frac{315\,x}{16\,y^6} + \frac{35}{4\,y^5} -
  \frac{135\,x}{8\,y^4} - \frac{21}{4\,y^3} +
  \frac{27\,x}{16\,y^2}
\right) \\&+
\left(
 \frac{12}{x^2}-20
 \right)
\phi_2(x)+
\frac{4\,\left( 1 - x^2 \right) }{x}
\phi_2'(x)
\\&
-\int_x^1 dy\,
\phi_2(y)
\left(
  \frac{315\,x}{16\,y^4}-\frac{7}{16\,x^3} + \frac{35}{4\,y^3} -
  \frac{135\,x}{8\,y^2} \right. \\& \left.
  - \frac{21}{4\,y}- \frac{15}{8\,x}
\right)
+
\frac{48}{3} \, \phi_4(x)  \\&
-
\int_x^1 dy
\, \phi_4(y)
\left(
\frac{45}{8\,x} + \frac{315\,x}{16\,y^2} + \frac{35}{4\,y} + \frac{15\,y}{4\,x^2} +
  \frac{35\,y^2}{16\,x^3}
\right) \;.
\end{split}
\label{Q4final}
\end{equation}
In principle, the derivation
of the analogous expressions for forward-like functions
$Q_{2\nu}$
with
$\nu \ge 3$
is straightforward. Unfortunately, the corresponding expressions
turn out to be too bulky.

\section{$D$-term and the divergencies of generalized form factors}
\label{Sec_toy}

In the framework of the dual parametrization of GPDs the small-$\xi$
behavior of the imaginary part of DVCS amplitude
(\ref{ImA_Dual})
inherits the singular behavior of the GPD quintessence function
\footnote{Below for simplicity we set the variable $t$ to zero. The generalization of our analysis for
$t \ne 0$ is straightforward.}
$N(x)$
for small $x$.
Assuming that
$N(x) \sim \frac{1}{x^{\alpha}}$
for
$x \sim 0$,
one can establish the following asymptotic behavior of the imaginary part
of the leading order DVCS amplitude (\ref{ImA_Dual})
\cite{Polyakov:2002wz}:
\begin{equation}
\im A(\xi) \sim
\frac{2^{\alpha+1}}{\xi^\alpha}
\frac{\Gamma(\frac{1}{2}) \, \Gamma(\alpha+\frac{1}{2})}
{\Gamma(\alpha+1)}\,.
\label{ImA_as_small_xi}
\end{equation}
It is usually supposed, that  the index
$\alpha$
of the leading singular power-like asymptotic behavior of the imaginary part
of DVCS amplitude for small
$\xi$
is determined by
the singular behavior of the corresponding forward
singlet quark distributions measured in DIS experiments:
$q_+(x) \sim \frac{1}{x^\alpha}$
for
$x \sim 0$
($0<\alpha<2$, $\alpha \ne 1$).
In the phenomenological applications of the dual parametrization
\cite{Guzey:2005ec,Guzey:2006xi}
it was tacitly assumed that the the leading small-$\xi$
singular behavior of the imaginary part of DVCS amplitude is entirely determined
by the small-$x$
behavior of the forward-like function
$Q_0(x)$:
\begin{equation}
q(x) \sim \frac{1}{x^\alpha} \ \ \ \rightarrow \ \ \ Q_0(x)
\sim \frac{ \alpha+\frac{1}{2}}{  \alpha+1}  \frac{1}{x^\alpha}\,.
\label{Q0_small_x_As}
\end{equation}
The functions
$Q_{2 \nu}(x)$
with high index
$\nu$
were supposed to be appropriately suppressed for small $x$
thus making no influence on the small-$\xi$
singular behavior of
$\im A(\xi)$.
However, the corresponding assumption is too restrictive and
by no means motivated from the physical point of view.
The forward-like functions
$Q_{2 \nu}(x)$
with
$\nu \ge 1$,
in principle, may contribute to the coefficient at the leading singular
power of
$\xi$
in the asymptotic expansion of
$\im A(\xi)$
for small
$\xi$.
For this one has to assume the following leading small-$x$
behavior of
$Q_{2 \nu}(x)$
with
$\nu \ge 1$:
\begin{equation}
Q_{2 \nu}(x) \sim \frac{1}{x^{2 \nu + \alpha}},
\label{Q2nu_as}
\end{equation}
where
$0 < \alpha <2$ ($\alpha \ne 1$).
A problem that immediately arises in this case is that the asymptotic
behavior
(\ref{Q2nu_as})
of
$Q_{2 \nu}(x)$
with
$\nu \ge 1$
leads to the divergencies of the generalized form factors
$B_{2 \nu-1 \;\; 0}$:
\be
B_{2 \nu-1 \;\; 0}= \int_0^1 \frac{dx}{x} \, x^{2 \nu} Q_{2 \nu}(x)\,.
\label{Bnl_with_sing}
\ee
According to
(\ref{Bnl_int}),
the generalized form factors
(\ref{Bnl_with_sing})
are the lowest order Mellin moments of the forward-like functions
$Q_{2 \nu}$
with
$\nu >0$
relevant for the calculation of GPDs in the framework of the dual
parametrization.
The crucial point is that
as it can be easily seen from
(\ref{DVCSampl_added}),
in the calculation of the leading order DVCS
amplitude the potentially divergent generalized form factors
(\ref{Bnl_with_sing})
contribute only into the $D$-form factor
(\ref{D_form_factor_NQ}).

Thus, we have to specify a suitable regularization of a
possible non-integrable singularity at
$x=0$
in (\ref{Bnl_with_sing}).
We restrict our following analysis to  the case in which the
contributions
$x^{2 \nu} Q_{2 \nu}(x)$
of the forward-like functions into GPD quintessence function
$N(x)$
as well as
$N(x)$
itself belong to the class of functions with power like behavior for
$x \sim 0$,
which can be presented as the
{\emph{finite}} sums of singular terms
\cite{Gelfand}:
\be
x^{2 \nu} Q_{2 \nu}(x), \; N(x) \in
\left\{ F: \; \;
F(x)=
\sum_{r=1}^R
\frac{1}{x^{\alpha_{r}}} f_{r}(x)
\right\}
\,. \nonumber \\
\label{Gelfand_Class}
\ee
Here
$f_{r}(x)$
with
$r=1,\,...R$
stand for arbitrary functions of
$x$
that are infinitely differentiable in the vicinity of
$x=0$.
It is also supposed that
$f_{r}(x)$
have zeroes of a sufficiently high order for
$x=1$.
We assume that
$0<\alpha_{r}<2$ ($\alpha_{r} \ne 1$)
so that a non integrable  singularity at
$x=0$
manifests itself only in the computation of the lowest order Mellin
moments
(\ref{Bnl_with_sing})
of
$Q_{2 \nu}(x)$
($\nu \ge 1$)
relevant for the calculation of GPD in the dual parametrization.
All higher order Mellin moments
$B_{n \;\; n+1-2 \nu}$
(\ref{Bnl_int})
with
$n= 2 \nu+1, \, 2 \nu+3, \,...$
turn to be finite. The class of functions defined in
Eq.~(\ref{Gelfand_Class})
seems to be convenient for model building. In particular, the
 functions usually employed in fitting of PDFs belong to this class. Moreover, the
imaginary part of the DVCS amplitude
$\im A(\xi)$
computed using the GPD quintessence function
$N(x)$
from the class
(\ref{Gelfand_Class})
also belongs to this class.

In order to work out a method of handling the  divergencies
of the generalized form factors
(\ref{Bnl_with_sing})
one has
to consider the general analytic properties of the DVCS amplitude.
According to the analysis presented in
\cite{Dispersion,Anikin:2007yh,Diehl:2007jb,Diehl:2007proc}
the leading order DVCS amplitude is considered to be a holomorphic function
of the variable
$\omega=\frac{1}{\xi}$.
The real part of the amplitude can be expressed through its imaginary
part with the help of a single variable dispersion relation in
$\omega$
for fixed value of
$t$.
%
In the singlet case this dispersion relation require one subtraction.
The subtraction constant is given by the value of the amplitude at
the non-physical point
$\omega=0\; (\xi = \infty)$.
It is known to be fixed by the
$D$
term and equals
$4 D^q$,
where
$D^q$
stands for the $D$-form factor
(\ref{Dff}).
After one switches back to the variable
$\xi$,
the once subtracted dispersion relation for the DVCS amplitude
(\ref{DVCS_ampl})
reads
\be
&&
A(\xi)= 4 D^q
\nonumber \\&&
+\frac{1}{\pi} \int_0^1 d \xi'
\left(
\frac{1}{\xi-\xi'-i \epsilon}-
\frac{1}{\xi+\xi'-i \epsilon}
\right)
\im A(\xi'-i \epsilon)\,.
\nonumber \\
\label{Disp_rel}
\ee
In general, the subtraction constant in a dispersion relation presents
an independent quantity, which cannot be fixed just with help of the information
on the discontinuities of the amplitude. Thus, in order to determine the
value of the subtraction constant one has to attain certain additional
information on the amplitude under consideration.

The problem of  fixing the subtraction constants in the dispersion relation for the
amplitude solely in terms of the absorptive part of the amplitude is addressed in the theory
of the analytic $S$-matrix. The key for this issue is provided by assuming the so-called
analyticity of the second kind that is the analyticity of the amplitude in the complex
angular momentum plane
\cite{Alfaro_red_book}.
The consideration of the analytical properties of the scattering amplitude
in the complex angular momentum plane is usually associated with the partial wave
expansion of the amplitude \cite{Col}.
However, as pointed out in
\cite{Xuri}
(see also discussion in
\cite{Alfaro_red_book}),
this type of analysis can equally be applied
for the expansions of the amplitude other than the partial wave expansion in particular
for the expansion of the amplitude into the ordinary power series.

This very logics was employed in
\cite{Mueller:2005ed,Kumericki:2007sa,Kumericki:2008di}
where it was suggested  to fix the subtraction constant
in the dispersion relation
(\ref{Disp_rel})
in terms of the imaginary part of the DVCS amplitude assuming analyticity properties
in $j$
of the specific combinations of coefficients
$h_{2 \nu}^{(2 \nu+j)}$
at powers of
$\xi$
of the Mellin moments of GPD
(see eq.~(\ref{Mellin_M}) for the definition).



For this following the line of analysis of
\cite{Mueller:2005ed,Kumericki:2007sa,Kumericki:2008di}
let us consider  the family of GPD sum rules for
the coefficients at powers of
$\xi$
of the Mellin moments of GPD
$h_{2 \nu}^{(2 \nu+j)}$.
Employing the dispersion relation
(\ref{Disp_rel})
together with
(\ref{DVCS_ampl})
one can establish the GPD sum rule
\cite{Dispersion,Anikin:2007yh,Diehl:2007jb}
\be
\int_0^1 dx
\left(
\frac{1}{\xi-x}-
\frac{1}{\xi+x}
\right)
\left[
H_+(x,\xi)-H_+(x,x)
\right]= 4D^q\,. \nonumber \\
\label{GPD_sum_rules}
\ee
Here the principle value prescription is dropped since the singularity at
$x=\xi$
turns to be integrable. Expanding the convolution kernel in
(\ref{GPD_sum_rules})
in powers of
$\frac{1}{\xi}$
and employing the polynomiality property
(\ref{Mellin_M})
of the Mellin moments of GPD one derives  a family of sum rules for the
coefficients at powers of
$\xi$
of Mellin moments of GPD:
\be
\sum_{\nu=1 }^\infty h_{2 \nu}^{(2 \nu+j)}=
\int_0^1 dx \, x^j
\left[
H_+(x,x)-H_+(x,0)\right]\,, \nonumber \\
\label{SR_GPD_h}
\ee
with $j=1,\,3,\,...\,$. Note that under the usual
assumptions on the small-$x$ asymptotic behavior:
\be
H_+(x,x)-H_+(x,0) \sim \frac{1}{x^\alpha} \ \ \text{with} \ \ \alpha<2 \ \ \ (\alpha \ne 1)
\nonumber \\
\label{Regge_Hxx}
\ee
the integral in (\ref{SR_GPD_h}) actually converges for odd positive $j$.

The  substantial step is to define the function
\be
\Phi(j) =\int_0^1 dx \, x^j
\left[
H_+(x,x)-H_+(x,0)\right] \nonumber \\
\equiv \frac{1}{\pi} \int_0^1 dx \, x^j
\left[
\im A(x)-\im A_{DIS}(x)\right]\,.
\label{Def_Phi_j}
\ee
Once the function
$\Phi(j)$
is analytically continued to $j=-1$ it
provides the desired relation for the subtraction constant in the dispersion relation
(\ref{Disp_rel})
\cite{Kumericki:2007sa}:
\be
&&
2 D^q=
\sum_{\nu=1 }^\infty h_{2 \nu}^{(2 \nu -1)}
=\Phi(j=-1) \,.
\label{Reg_atar}
\ee
In order to take advantage of the analyticity of $\Phi(j)$ in
the complex
$j$
plane we have to specify our assumptions for
the analytic properties of the function
$H_+(x,x)-H_+(x,0) \equiv \frac{1}{\pi} \left[ \im A(x)-\im A_{DIS}(x) \right]$
in the
$x$
plane.
We suppose  the Regge-like asymptotic behavior
(\ref{Regge_Hxx})
of
$\im A(x)-\im A_{DIS}(x) $
for small-$x$
which can be established experimentally.
In our present analysis in order to obtain simple analytical properties
of
$\Phi(j)$
in the complex
$j$
plane
we assume that
$\im A(x)-\im A_{DIS}(x)$
belongs to the class of functions defined in
Eq.~(\ref{Gelfand_Class})%
\footnote{The generalization for $\im A(x)-\im A_{DIS}(x)$, which belong to the more intricate classes
like e.g.
$
F: \,F(x)=
\sum_{r=1}^R
\frac{1}{x^{\alpha_{r}}} (\log{x})^{\beta_r} f_{r}(x)$
is straightforward. }.
Under this assumption the function
$\Phi(j)$
turns to be a meromorphic function of
$j$ for $\re j >-1-\epsilon$ ($\epsilon>0$).
$\Phi (j)$ has just a finite number of simple poles in the complex plane $j$ for $\re j >-1$.
In this case the analytic continuation of $\Phi (j)$ to $j=-1$ exists and is obviously unique.
It is provided by the following explicit expression for the regularization of the divergent integrals in
(\ref{Def_Phi_j}):
\be
&&
{\rm{Reg}}
\int_{0}^1 dx \,
 \frac{f(x)}{x^{1+\alpha}} \equiv
\int_{(0)}^1 dx \,
 \frac{f(x)}{x^{1+\alpha}} \nonumber \\&&=
\int_{0}^1 dx \,
 \frac{1}{x^{1+\alpha}}
\left[f(x)-f(0)-x f'(0) \right]
\nonumber \\&&
 -f(0)\frac{1}{\alpha}  -f'(0)\frac{1}{\alpha-1}\,.
 \label{Areg} 
\ee
Clearly this is what is known as the analytic regularization of integrals
\cite{Gelfand}.

Thus, under the specified above analyticity assumptions
(which, however,  can be violated, as we discuss below), the subtraction constant
in the dispersion relation
(\ref{Disp_rel})
can be fixed in terms of the imaginary part of the DVCS amplitude
according to
\cite{Kumericki:2008di}
\be
2 D^q= \int_{(0)}^1 dx \frac{1}{x}
\left[
H_+(x,x)-H_+(x,0)
\right]\,.
\label{subtraction_constant_anlytic}
\ee

Now we discuss how these general considerations apply for
the case of the dual parametrization of GPDs. We argue that the implementation of
analyticity of the combinations
(\ref{GPD_sum_rules})
of coefficients of the Mellin moments of GPDs suggests the use of analytic
regularization for the potentially divergent generalized form factors
$B_{2 \nu-1 \; \;0}$
(\ref{Bnl_with_sing}).

Let us consider the sum rule
(\ref{subtraction_constant_anlytic})
for the $D$- form factor in the framework of the dual parametrization of GPDs.
The explicit expressions for
$H(x,x)$
and
$H(x,0)$
through the forward like functions in the framework of  the dual parametrization read
\be
H(x,0)=
Q_0(x)+
\frac{\sqrt{x}}{2}
\int_x^1
\frac{dy}{y^{\frac{3}{2}}}
Q_0(y)
\ee
and
\be
H(x, x )= \frac{2}{\pi}
\int_{\frac{1- \sqrt{1-x^2}}{x}}^1
\frac{dy}{y}
N(y)
\frac{1}{\sqrt{\frac{2y}{x}-y^2-1}}\,.
\ee
We assume that both functions
$N(y)$
and
$Q_0(y)$
belong to the class (\ref{Gelfand_Class}).
The leading singular behavior for
$y \sim 0$
is
$Q_0(y), \; N(y) \sim \frac{1}{y^\alpha}$ with $\alpha<2$.
It is straightforward  to check that interchanging the order of integration
\be
&&
\frac{2}{\pi}
\int_{(0)}^1 \frac{dx}{x}
 \int_{\frac{1- \sqrt{1-x^2}}{x}}^1 \frac{dy}{y} N(y)
\frac{1}{\sqrt{\frac{2y}{x}-y^2-1}}
 \rightarrow
\nonumber \\ &&
\frac{2}{\pi}  \int_{(0)}^1 \frac{dy}{y} N(y) \int_0^{\frac{2y}{1+y^2}} \frac{dx}{x} \frac{1}{\sqrt{\frac{2y}{x}-y^2-1}}
\ee
is a rigorously defined operation for the class of functions under consideration.
The same is true for
\be
&&
\int_{(0)}^1 \frac{dx}{x}  \frac{\sqrt{x}}{2}\int_x^1 \frac{dy}{y^{\frac{3}{2}}} Q_0(y)  \rightarrow
\int_{(0)}^1 \frac{dy}{y^{\frac{3}{2}}} Q_0(y)  \int_0^y \frac{dx}{2 \sqrt{x}}\,.
\nonumber \\ &&
\ee
In this way one arrives to the analytically regularized version of
the expression
(\ref{D_form_factor_NQ})
for the $D$ form factor%
\footnote{In
\cite{Polyakov:2007rw}
an alternative regularization prescription was suggested.
The two regularizations obviously  differ just by a finite constant
and thus are equivalent up to a {\emph{finite}} $D$-term contribution.
}:
\begin{equation}
\begin{split}
D^q=&
\int_0^1 \frac{dy}{y}N(y)
\left(
\frac{1}{\sqrt{1+x^2}}
-1
\right)+ \\&
\int_{(0)}^1 \frac{dy}{y}
\left[
N(y)-Q_0(y)
\right]
\,.
\end{split}
\label{D_form_factor_NQ_reg}
\end{equation}
The integral in the first term of
(\ref{D_form_factor_NQ_reg})
converges under our assumptions on the small-$y$
behavior of $N(y)$ so it does not require analytic regularization.
The second term is the pure contribution of the problematic form factors  $B_{2 \nu-1 \; 0}$
since
$$
\int_{(0)}^1 \frac{dy}{y}
\left[
N(y)-Q_0(y)
\right]= \int_{(0)}^1 dy  \sum_{\nu=1}^\infty   \frac{1}{y} y^{2 \nu} Q_{2 \nu}(y)\,.
$$
So we conclude that the analytic regularization of the integral
in the second term of
(\ref{D_form_factor_NQ_reg})
suggests the use of the analytic regularization for
the individual terms of the sum
\be
B_{2 \nu-1 \; 0}= \int_{(0)}^1   \frac{dx}{x} x^{2 \nu} Q_{2 \nu}(x)\,.
\ee



It is also  extremely instructive to check that under our assumption on
the analyticity of the combinations
(\ref{Def_Phi_j})
of coefficients of the Mellin moments GPDs in the framework of the dual parametrization
satisfy the so-called ``duality property'' suggested in
\cite{Kumericki:2008di}.
According to the ``duality property'', GPD in the central region
$0<x<\xi$
can be completely restored from its knowledge in the outer region
$\xi<x<1$.
The coefficients at powers of $\xi$ of $N$-th Mellin moment
($N=1,\,3,\,...$)
of GPD
$H_+(x,\xi)$
can be determined from the small
$\xi$
expansion
(\ref{small_xi_dual})
of
$H_+(x,\xi)$
in powers of
$\xi$
for
$x>\xi$
according to
\be
&&
\int_0^1 dx \, x^N H_+(x,\xi)=
\sum_{\nu=0}^{\frac{N+1}{2}} \xi^{2 \nu} h_{2 \nu}^{(N)}=
\nonumber \\ &&
\sum_{\nu=0}^{\frac{N-1}{2}} \xi^{2 \nu}  \int_{0}^1 dx\,x^N
H_+^{(\nu)}(x)+
\xi^{N+1} \int_{(0)}^1 dx\,x^N
H_+^{(\frac{2N+1}{2})}(x) \,, \nonumber \\
\label{Duality_property}
\ee
where
$H_+^{(\nu)}(x)= \frac{\partial^{2 \nu}}{\partial \xi^{2 \nu}}
H_+(x,\xi)_{\xi=0}$.
The potentially divergent integral for the coefficient at the highest
power of
$\xi$
is understood as analytically regularized.
Assuming that the forward like functions belong to the class
(\ref{Gelfand_Class})
and employing the available explicit expressions for $H_+^{(\nu)}(x)$ one checks that
the ``duality property'' (\ref{Duality_property})
holds in the framework of the dual parametrization of GPDs.
Thus all Mellin moments of GPD
can be entirely determined from the knowledge of GPD in the outer region
($x>\xi$).
Now, {\it e.g.} employing the standard inverse Mellin transform techniques one can
recover GPD in the whole region from its Mellin moments.
In this sense the GPD in the outer region
($x>\xi$) determines GPD in the whole region $0 \le x \le 1$

Let us stress once more that the possibility to fix the
$D$-form factor in terms of the amplitude strongly relies on the postulated
analyticity of the specific combination of the coefficients of Mellin moments of GPDs in the
Mellin space.
This analyticity assumption
seems to be alluring from the theoretical point of view
for modeling GPDs in the framework of the dual parametrization
since it enables to treat spin-$0$ exchange contributions on the equal footing
with that of higher spins. It also ensures the ``good'' analytic properties of
$\im A(\xi)$
(or equivalently forward-like functions).
However, there is still no final confidence on the validity of this analyticity assumption
(see discussion in
\cite{Kumericki:2007sa}).
For example, this kind of analyticity
can be
absent due to the contact interaction contribution
to the real part of the DVCS amplitude or due to
the so-called fixed pole singularity at
$j=-1$
(see e.g. \cite{Col}).
In
(\ref{Reg_atar})
these kind of singularities reveal themselves as terms proportional to a Kronecker
$\delta_{-1 j}$
which is {\emph{non analytic}} in $j$.

As it was pointed out in
\cite{ForwardLikeF_KS},  in the chiral quark model there is the explicit contact term contribution into the singlet pion GPD
required by the chiral symmetry
(see diagram~(3) in Fig.~5 of  Ref \cite{ForwardLikeF_KS})
which violates the suggested analyticity property.

Once the requirement of analyticity is lifted, the $D$-term may introduce
an independent contribution
into the real part of DVCS amplitude. In this case, the $D$-form factor
becomes an independent physical quantity to be fixed from the experiment.
Adding a supplementary $D$-term
$
\theta(1-\frac{x^2}{\xi^2}) \, \delta{D}\left(\frac{x}{\xi}\right)
$
with the Gegenbauer expansion
\begin{equation}
\delta{D}(z)=(1-z^2) \sum_{n=1 \atop \text{odd}}^\infty \delta{d}_n\,
C_n^{\frac{3}{2}}(z)
\label{D_tilde_term}
\end{equation}
to a GPD is equivalent to an introduction of the following {\emph{non analytic}}
contributions to the forward-like functions:
\be
x^{2 \nu} Q_{2 \nu}(x) \longrightarrow
x^{2 \nu} Q_{2 \nu}(x)+2 \delta{d}_{2 \nu-1} \, x \delta (x)\,.
\ee

Another possible source of the non-analytic contributions to the sum rule
(\ref{Reg_atar})
may be the existence of a fixed pole
(see \cite{Alfaro_red_book,Col})
at
$j=-1$
(i.e. angular momentum
$l=0$).
This type of contribution  was reasoned with the Regge
theory inspired argumentation for the case of forward Compton scattering amplitude in
\cite{Creutz:1968ds}
and revealed in the experimental measurements
\cite{Damashek:1969xj,Dominguez:1970wu}.
Nevertheless, according to
\cite{Cornwall:1970dp,Cornwall:1971pk,Brodsky:1971zh,Brodsky:1973hm}
via a subtracted sum rule the fixed pole contribution can be related to the
imaginary part of Compton amplitude. Once  generalized for the case of
DVCS  these considerations would also imply no independent $D$-term contribution into
DVCS amplitude.


\section{$Q_2(x)$
and
$Q_4(x)$
from the small
$\xi$
expansion
of
$H_{DD}(x,\xi)$}
\label{Sec_Q2Q4_for_double}

One of the most popular parametrizations of GPDs consists in the use of
the double distributions
\cite{Radyushkin:1997ki,Radyushkin:1998bz,RadDDandEvolution,Musatov:1999xp,Radyushkin:2000uy}.
In order to make this parametrization consistent with
the polynomiality condition in its full form one has to complete it
with the so-called $D$-term
\cite{Polyakov:1999gs}.
More technically, in this case the nucleon GPD
$H^q$
for the particular flavor
$q$
is parametrized in the following form%
\footnote{Note that since $t$-dependence plays no particular role
in our considerations we just set $t$ to zero. }:
\begin{equation}
  H^q(x,\xi,t=0)=H_{DD}^q(x,\xi)+ \theta(\xi-|x|)
  D^q \left(\frac{x}{\xi} \right),
  \label{2componentparam}
\end{equation}
where
$H_{DD}^q$
is obtained as a one dimensional section of a two-variable
double distribution
$F^q$:
\begin{equation}
H_{DD}^q(x,\xi)= \int_{-1}^1 d \beta \,
\int_{-1+| \beta|}^{1-|\beta|} d \alpha \,
\delta(x- \beta - \alpha \xi) \, F^q(\beta, \alpha).
\label{DDpart}
\end{equation}
According to the proposal of Radyushkin
\cite{RadDDandEvolution,Musatov:1999xp,Radyushkin:2000uy},
the following model for
the double distribution
$F^q$
is often used:
\begin{equation}
F^q(\beta, \alpha)= h(\beta, \alpha) q(\beta),
\label{DDansatz}
\end{equation}
where the profile function
$h(\beta, \alpha)$
is parameterized through the following favored Ansatz:
\begin{equation}
h^{(b)}(\beta, \alpha)=\frac{\Gamma(2b+2)}{2^{2b+1} \Gamma^2(b+1)}
\frac{[(1-|\beta|)^2-\alpha^2]^b}{(1-|\beta|)^{2b+1}}
\label{RadProfile}
\end{equation}
and
$q(\beta)$
($q(-\beta)=-\bar{q}(\beta)$)
stands for the phenomenological forward quark distribution.
The parameter
$b$
characterizes the strength of
$\xi$
dependence of the resulting GPD
$H^q(x,\xi)$.
The limiting case
$b \rightarrow \infty$
corresponds to
$\xi$
independent Ansatz
$H^q(x,\xi)=\theta(x) q(x)-\theta(-x)\bar{q}(-x)$,
considered {\it e.g.} in
\cite{Guichon:1998xv}.
The GPD model
(\ref{DDpart})
based on the Ansatz
(\ref{DDansatz}), (\ref{RadProfile})
for DD has been used in numerous phenomenological
applications. In particular, the Radyushkin model is implemented in the popular VGG
code (see \cite{VGG}).

In accordance with the standard expressions for the singlet combination of GPDs
(which is reduced to
$\,q_+(x) \equiv q(x)+\bar{q}(x)$
in the forward limit) we define:
\begin{equation}
 \begin{split}
  & H_{+ \,DD}(x, \xi) \equiv
  H^q_{DD}(x, \xi)-H^q_{DD}(-x, \xi)\\& =
  \int_0^1 d \beta
  \int_{-1+\beta}^{1-\beta}
  d \alpha
  \left\{
   \delta(x-\beta- \alpha \xi)-
    \delta(x+\beta- \alpha \xi)
  \right\} \\& \times
 h^{(b)}(\beta, \alpha)
 q_+(\beta)\,.
 \end{split}
 \label{H_DD_x>xi}
\end{equation}
We now compute the coefficients at powers of
$\xi$
of $N$-th ($N$- odd) Mellin moment of the quark singlet GPD
$H_{+\,DD} (x,\xi)$:
\be
  && \int_{0}^1dx \, x^N H_{+\,DD} (x,\xi) \nonumber \\&& =
  h_0^{(N) \,DD}+ h_2^{(N) \, DD}\xi^2+....h_{N-1}^{(N)\, DD}\xi^{N-1};
\nonumber \\&&  h_{N+1}^{(N) \,DD} \equiv 0 \,.
\ee
For this we perform the straightforward calculation:
\begin{equation}
\begin{split}
&  
\int_{0}^1 dx \, x^N
\int_0^1 d \beta
  \int_{-1+\beta}^{1-\beta}
  d \alpha
  \left\{
   \delta(x-\beta- \alpha \xi)-
    \delta(x+\beta- \alpha \xi)
  \right\} \\& \times
 h^{(b)}(\beta, \alpha)
 q_+(\beta)
 \\&
=\int_0^1 d \beta \,
\int_{-1+\beta}^{1-\beta} d \alpha \,
\sum_{k=0}^N  \xi^k C_N^k \, \alpha^k \beta^{N-k}
h^{(b)}(\beta, \alpha) q_+(\beta)\,,
\end{split}
\end{equation}
where
$C_N^k$ stand for the usual binomial coefficients ($C_N^k= \frac{N!}{(N-k)!k!}$).
Now using
\begin{equation}
\begin{split}
 & \int_{-1+\beta}^{1-\beta} d \alpha \, \alpha^k
h^{(b)}(\beta, \alpha) \,   \\& =
\left\{
\begin{array}{ll}
0 \ \ , & k  \ \ \ \text{odd}; \\
\frac{(k-1)!! \; \Gamma(b+\frac{3}{2})}{2^{\frac{k}{2}} \Gamma(b+\frac{k+3}{2})}(1-\beta)^k \ \ , & k  \ \ \
\text{even};
\end{array}
\right. \ \ \text{for} \ \  \beta \ge 0
\end{split}
\end{equation}
we obtain the following expression for the coefficients at powers of $\xi$ of $N$-th Mellin
moment of the singlet GPD in the framework of the double distribution parametrization:
\be
&& h^{(N) \, DD}_{2 \nu} \nonumber \\&&
=
\int_0^1 d \beta\, \beta^{N-2 \nu} (1-\beta)^{2 \nu} C_N^{2 \nu}
\frac{(2 \nu-1)!! \, \Gamma(b+\frac{3}{2})}{2^{\nu} \Gamma(b+\frac{2 \nu+3}{2})}
\, q_+(\beta) \nonumber \\&&
\label{hk_double}
\ee
for $ \nu=0, \,1, \, ... \,\frac{N-1}{2}$.


Our present goal consists in deriving the expressions
for the forward-like functions, which  allow us to rewrite Radyushkin double
distribution Ansatz through the dual parametrization. For this we employ the expansion
(\ref{Expansion_start})
of
$H_{+\,DD}(x,\xi)$
in powers of small
$\xi$
for
$x>\xi$.
Such an expansion was originally constructed in
\cite{RadDDandEvolution,Musatov:1999xp}:
\begin{equation}
\begin{split}
& H_{+\,DD}(x, \xi) \\&
=q_+(x)+
\frac{\xi^2}{2!}
\frac{1}{(2b+3)}
\frac{\partial^2}{\partial x^2}
\left(
q_+(x)
(1-x)^2
\right) \\& +
\frac{\xi^4}{4!}
\frac{3 }{(2b+3)(2b+5) }
\frac{\partial^4}{\partial x^4}
\left(
q_+(x)
(1-x)^4
\right)+O(\xi^6)  \\& \equiv
\phi_{0 \, DD}^{(b)}(x)+
\xi^2
\phi_{2 \, DD}^{(b)}(x)+
\xi^4
\phi_{4 \, DD}^{(b)}(x)+ O(\xi^6).
\end{split}
\label{Small_xi_Rad_Exp}
\end{equation}
As it was pointed out in
\cite{Musatov:1999xp},
if the singular behavior
$q_+(x) \sim \frac{1}{x^\alpha}$
is assumed,
the relative suppression of
the consecutive corrections to the expansion
(\ref{Small_xi_Rad_Exp})
is not
$O(\xi^{2 \nu})$
but rather
$O(\xi^{2 \nu}/x^{2 \nu})$.
As a consequence, this expansion is efficient only for
$x \gg \xi$.
Interestingly, the expansion
(\ref{Small_xi_Rad_Exp})
produces the correct values
(\ref{hk_double})
of the coefficients
$h^{(N) \, DD}_{2 \nu}$:
\be
\int_{0}^1 dx
\, x^N  \phi_{2 \nu \, DD}^{(b)}(x)=
h^{(N) \, DD}_{2 \nu}, \ \ \ \nu=0,... \, \frac{N-1}{2}\,.
\nonumber \\
\label{hdd_nonsing}
\ee
In contrast, the expression for the coefficient at the highest power of
$\xi$
of the $N$-th Mellin moment turns out to be singular. Assuming the ``good''
analytical properties of Mellin moments of
$H_{+\, DD}$
in Mellin space one can treat this singularity
exactly in the same way as described in
sect.~\ref{Sec_toy}:
\be
\int_{(0)}^1 dx
\, x^N  \phi_{N+1 \, DD}^{(b)}(x)=
h^{(N) \, DD}_{N+1}\,.
\label{hdd_sing}
\ee
One can check that
the use of analytic regularization
(\ref{Areg})
leads to
the correct values
$h^{(N) \, DD}_{N+1}=0$. Thus from
(\ref{hdd_nonsing}), (\ref{hdd_sing})
we conclude that GPD in the framework of double distribution
parametrization without $D$-term satisfies the ``duality property'' of Ref.~\cite{Kumericki:2008di}.
Namely, the GPD in the central region
$0<x<\xi$
can be completely restored from its knowledge in the outer region
$\xi<x<1$.

Employing the expansions
(\ref{Small_xi_Rad_Exp})
and
(\ref{small_xi_dual})
we establish the correspondence between the double distribution Ansatz and
the dual parametrization.
With the help of the general result
(\ref{Q2final})
the following expression for the forward-like function
$Q_2(x)$
can be derived:
\begin{equation}
Q_2(x)=f_{Q_2}^{(b)}(x)+ \int_x^1 dy \, K_{Q_2}^{(b)}(x,y) \, q_+(y)\,,
\label{Q2DD}
\end{equation}
where
\begin{equation}
\begin{split}
&f_{Q_2}^{(b)}(x)=
\left(
2\,\left( 1 - x^2 \right)  +
  \frac{9 - 30\,x + 29\,x^2}{2\,\left( 3 + 2\,b \right) }
\right) \frac{1}{x^2} \,q_+(x) \\&
+ \left(1 + x+ \frac{3 - 11\,x}{3 + 2\,b}
\right) \frac{(1-x)}{x} \,
q_+'(x)+
\frac{2\,{\left( 1 - x \right) }^2}{3 + 2\,b} \,
q_+''(x)
\end{split}
\label{Q2_main}
\end{equation}
and the convolution kernel
$K_{Q_2}^{(b)}(x,y)$
is given by
\be
&&
K_{Q_2}^{(b)}(x,y) \nonumber \\ && =
\left(
-6\,y - \frac{3\,{\left( 1 - y \right) }^2\,\left( 15\,x + 2\,y \right) }
   {3 + 2\,b} - 5\,x\,\left( 3 - y^2 \right)
\right) \frac{1}{4y^4} \,.
\nonumber \\ &&
\ee
Note that if  power-like small-$x$
behavior is assumed for
$q_+(x)$:
\begin{equation}
q_+(x) \sim \frac{1}{x^\alpha}\,, \ \ \ \alpha<2\,
\label{ReggeLikeq}
\end{equation}
the small-$x$ asymptotic behavior of
$Q_2(x)$ is:
$Q_2(x) \sim \frac{1}{x^{\alpha+2}}$.
The divergency occurring in the
generalized form factor
$B_{1 \,0}$
should be treated as described in
sect.~\ref{Sec_toy}.
Since we started from
$H_{+\,DD}$
(\ref{H_DD_x>xi})
with no
$D$-term included it is important to check that no $D$-term
is generated by the described above reparametrization procedure.
%

The forward-like function
$Q_4(x)$
can be expressed in a completely analogous way with the
help of
(\ref{Q4final}):
\begin{equation}
Q_4(x)=f_{Q_4}^{(b)}(x)+ \int_x^1 dy \, K_{Q_4}^{(b)}(x,y) \, q_+(y).
\label{Q4DD}
\end{equation}
The explicit expressions for
$f_{Q_4}^{(b)}(x)$
and the convolution kernel
$K_{Q_4}^{(b)}(x,y)$
are presented in the Appendix~\ref{Expl_Q4_DD}.
Unfortunately, the expressions for
the forward-like functions
$Q_{2\nu}$
with
$\nu \ge 3$
turn out to
be very bulky. However, as it is shown in the next sect. for the case
of Radyushkin double distribution parametrization several first
forward-like functions provide dominant contribution into
GPD quintessence function
$N(x)$ for small-$x$.

In order to compare the predictions of  the double distribution parametrization and
the dual parametrization of GPDs it is extremely instructive to consider the asymptotic
behavior of the imaginary part of the DVCS amplitude at small values of
$\xi$
calculated in these two models.
Let us assume the power-like asymptotic behavior
(\ref{ReggeLikeq})
for
$q_+(x)$.
From the double distribution parametrization one can easily derive the leading asymptotic behavior of the
imaginary part of the DVCS amplitude at small
$\xi$:
\be
&&
\text{Im}A_{DD}(\xi)=
\frac{\pi}{\xi}
\int_0^{\frac{2 \xi}{1+\xi}} dx \, h^{(b)} \left(x, \frac{\xi-x}{\xi}\right) \, q_+(x) 
\nonumber \\&&
\sim
\frac{2^{2b+1-\alpha}}{\xi^\alpha}
\frac{\Gamma(\frac{1}{2}) \Gamma(b+ \frac{3}{2}) \Gamma(1+b-\alpha)}{\Gamma(2+2b-\alpha)}
\label{ImA_DD_small_xi_asymp}
\ee
Let us now consider the dual parametrization.
The contribution of the particular forward-like function
$Q_{2 \nu}$
into the imaginary part of the amplitude reads
\begin{equation}
\im A^{(\nu)}(\xi)= 
2 \int_{\frac{1- \sqrt{1-\xi^2}}{\xi}}^1
\frac{dx}{x} \, x^{2 \nu}
Q_{2 \nu}(x)
\left[
\frac{1}{\sqrt{\frac{2x}{\xi}-x^2-1}}
\right]\,; 
\label{ImA_nuDual}
\end{equation}
Regge-like behavior
(\ref{ReggeLikeq})
of the forward quark distribution
leads to asymptotic behavior
(\ref{Q0_small_x_As})
of the corresponding forward-like function
$Q_0$
for small
$x$.
For the leading asymptotic behavior of its contribution into
the imaginary part of the amplitude for
$\xi \sim 0$
we obtain
(\ref{ImA_as_small_xi}):
\begin{equation}
  \text{Im}A^{(0)}(\xi) \sim \frac{2^{\alpha+1}}{\xi^\alpha}
  \frac{\Gamma(\frac{1}{2}) \Gamma(\alpha+ \frac{3}{2})}{\Gamma(\alpha+2)}
\end{equation}
Clearly, for
$\alpha=b$
the coefficients in front of leading singular term of
$\text{Im}A_{DD}(\xi)$
(\ref{ImA_DD_small_xi_asymp})
and the
$Q_0(x)$
contribution into the imaginary part
$\text{Im}A^{(0)}(\xi)$
coincide%
\footnote{It is interesting to note that the factorized Ansatz with the correlation
$b=\alpha$ between the profile function parameter $b$ and the power characterizing the small-$x$ behavior
of the forward distribution was first addressed in \cite{Musatov:1999xp}. It was found to correspond to the
case when the Gegenbauer moments of GPD are $\xi$-independent. This very assumption was the starting point
for the model for GPDs constructed in \cite{Shuvaev:1999ce}.
}
.
This fact is by no doubts responsible for the
approximate equality
\cite{Guzey:2006xi}
of imaginary the parts of
the singlet quark  GPD contributions into
the DVCS amplitude calculated in the minimal dual model
(with only $Q_0$ kept)
to that calculated in the framework of the double distribution
parametrization with $b=1$ for small values of $\xi$.
Additionally  taking into account  the contributions of
$Q_2$
(\ref{Q2DD})
and $Q_4$
(\ref{Q4DD})
we derive the following asymptotic behavior for
the imaginary part of the amplitude
$ \im A (\xi)$:
\begin{equation}
\begin{split}
& \im A^{(0)}(\xi)+\im A^{(1)}(\xi)+\im A^{(2)}(\xi) \\ & \sim
\frac{2^{\alpha+1}}{\xi^\alpha}
  \frac{\Gamma(\frac{1}{2}) \Gamma(\alpha+ \frac{3}{2})}{\Gamma(\alpha+2)}
\left(
1+
\frac{\left(  \alpha-b  \right)}{\left( 3 + 2\,b \right)}
\frac{2\,\alpha \,\left( 1 + \alpha  \right) \,\left( \frac{5}{2} + \alpha  \right) \,
     }{ \,\left( 2 + \alpha  \right) \,
    \left( 3 + \alpha  \right) }
\right. \\&
+ \left.
\frac{  \left(  \alpha-b  \right) \left( \alpha -b+1 \right)}
{\left( 3 + 2\,b \right) \,\left( 5 + 2\,b \right)}
\frac{2 \,\alpha \,
    \left( 1 + \alpha  \right) \,\left( \frac{3}{2} + \alpha  \right) \,
    \left( \frac{9}{2} + \alpha  \right) }{ \left( 4 + \alpha  \right) \,\left( 5 + \alpha  \right) }
\right)\,.
\label{Small_xi_as_Im_A}
\end{split}
\end{equation}
Note, that the consideration of contributions of additional forward-like
functions does not spoil the coincidence of asymptotic behavior of
$\im A(\xi)$
calculated from the dual and the double distribution parameterizations
with the same input for
$b=\alpha$.
Moreover, for
$b=\alpha+1$
the contributions of
$Q_0$ and $Q_2$ result  in small-$\xi$
asymptotic behavior
of $\im A(\xi)$
coinciding with that of
$\im A_{DD}(\xi)$.
Analogously for
$b=\alpha+2$
one just needs to take account of the contributions of
$Q_0$, $Q_2$
and
$Q_4$.
In general
for $b=\alpha+M$, $M>0$, integer,  it suffices to take account
of a finite number of forward-like functions
$Q_{2 \nu}$
with $\nu \le M$
obtained using the reparametrization procedure
in order to reproduce the leading small-$\xi$
asymptotic behavior
(\ref{ImA_DD_small_xi_asymp})
of
$\im A_{DD}(\xi)$.

Let us also consider the asymptotic behavior of the imaginary part of the
amplitude for
$\xi \sim 1$.
Assuming that
\begin{equation}
q_+(x) \sim (1-x)^\gamma\,, \ \ \ \gamma>0\, \ \ \ \text{for} \ \ x \sim
1
\label{q_sim_1}
\end{equation}
the leading asymptotic behavior
for
$\xi \sim 1$
of the
imaginary part of the amplitude calculated in the
framework of Radyushkin double distribution parametrization
is given by
\be
\text{Im}A_{DD}(\xi)
%
\sim
\, 2^{b}
\sqrt{\pi} \, \frac{\Gamma \left(  \frac{3}{2}+b \right)
\Gamma(\gamma-b)}{\Gamma(1+\gamma)}(1-\xi)^b\,.
\label{asympIm_DD_xi_1}
\ee

For the dual parametrization with the same input
(\ref{q_sim_1})
the leading asymptotic
behavior of
$\im A^{(0)}(\xi)$
for $\xi \sim 1$
is given by:
\begin{equation}
  \text{Im}A^{(0)}(\xi) \sim
2^{\frac{\gamma}{2}}
  \sqrt{\pi} \frac{\Gamma \left(
\frac{1+\gamma}{2}
  \right)}{\Gamma \left( \frac{\gamma}{2}+1 \right)}
  (1-\xi)^{\frac{\gamma}{2}}.
  \label{asympIm_dual_xi_1}
\end{equation}
Contrary to the case of small-$\xi$
asymptotic behavior of
$\im A(\xi)$,
for
$\xi \sim 1$
it is not possible
to make the dual and double distribution parametrizations with the
same input
(\ref{q_sim_1})
result in the
identical asymptotic behavior of
$\im A(\xi)$
simply by adjusting the value of the
parameter
$b$.

The two models lead to the substantially different asymptotic behavior for
the imaginary part of the DVCS amplitude
for
$\xi \sim 1$.
In order to reproduce the asymptotic behavior
(\ref{asympIm_DD_xi_1})
in the framework of the dual parametrization
one has to take into account the contribution of the whole set of
the forward-like functions $Q_{2 \nu}$ computed with the
help of the described above reparametrization procedure.

Moreover, it is important to stress that the asymptotic behavior
(\ref{asympIm_DD_xi_1})
of
$\im A_{DD}(\xi)$
for
$\xi \sim 1$
calculated in the double distribution parametrization
is determined by the value of the
parameter $b$, which entirely specifies the power-like behavior of
$\im A(\xi)$
in powers
of
$(1-\xi)$
for an arbitrary input forward quark distribution.
The same parameter $b$ also determines the
coefficient at the leading singular term of
$\im A_{DD}(\xi)$
for
$\xi \sim 0$
(\ref{ImA_DD_small_xi_asymp}).
Thus
in the framework of the Radyushkin double distribution Ansatz
the two  types of  asymptotic behavior of
$\im A(\xi)$
(namely, for $\xi \sim 0$ and $\xi \sim 1$ )
can not be described independently.
This can be seen as the important drawback of this Ansatz
making it not enough flexible for the description of the whole set of the
DVCS data.

On the contrary, the asymptotic behavior of
$\im A (\xi)$
for
$\xi \sim 1$
calculated in the framework of the dual parametrization
turns out to be sensitive to the
$x \sim 1$
behavior of the GPD quintessence function $N(x)$ for $x \sim 1$:
\be
N(x) \sim O((1-x)^\gamma)   \ \ \ \longrightarrow \ \ \ \im A(\xi) \sim O((1-\xi)^{\frac{\gamma}{2}})\,.
\nonumber \\
\ee
As a result, in the framework of the dual parametrization, one can model the
$\xi \sim 1$
asymptotic behavior of
$\im A (\xi)$
independently from that for $\xi \sim 0$.
Thus, the dual parametrization of GPDs is more
flexible and hence may turn to be more convenient for the description of the DVCS
data for both large and small values of
$\xi$.

\section{Numerical results for
$Q_0(x)$,
$Q_2(x)$
and
$Q_4(x)$
calculated from the double distribution parametrization
of nucleon GPD $H(x, \xi)$}

It is extremely instructive to compare our
results for the forward-like functions
$Q_0$, $Q_2$, $Q_4$
(\ref{Q0final}), (\ref{Q2DD}), (\ref{Q4DD})
calculated from the double distribution model
to the general form of GPD quintessence function
$N(x)$
(\ref{Quintes}),
which
can be recovered from the known
imaginary part of the amplitude
$\im A(\xi)$
with the help of Abel tomography
method \cite{Tomography,Polyakov:2007rw,Moiseeva:2008qd}
\footnote{Note that the Abel transformation was first employed in connection with
GPDs in \cite{Musatov:1999xp}
for constructing a model for DD related to PDF by Abel integral equation.
More general inverse Radon transformation was used in
\cite{Teryaev:2001qm} to relate GPDs and GDAs to DDs.}:
\begin{equation}
\begin{split}
& N(x) =
\frac{1}{2\pi}
\frac{\sqrt{2x}(1+x)}{ \sqrt{1+x^2}} \, \im A
\left(
\frac{2x}{1+x^2}
\right) \\&
-\frac{1}{2\pi} 
\frac{x(1-x^2)}{(1+x^2)^{\frac{3}{2}}}
\int_{\frac{2x}{1+x^2}}^1 d \xi \,
\frac{1}{(\xi-\frac{2x}{1+x^2})^{\frac{3}{2}}}
\left\{
\frac{1}{\sqrt{\xi}} \, \im A(\xi) \right. \\&
\left.
-
\sqrt{\frac{1+x^2}{2x}} \,
\im A
\left(
\frac{2x}{1+x^2}
\right)
\right\}\,.
\end{split}
\label{NfunctionA}
\end{equation}
Using the double distribution parametrization result for the imaginary
part of the amplitude
\begin{equation}
  \text{Im} A_{DD}(\xi)=\frac{\pi}{\xi} \int_0^{\frac{2 \xi}{1+\xi}}
  d \beta \, h^{(b)} \left( \beta, \, \frac{\xi-\beta}{\xi} \right) \,
  q_+(\beta)
\end{equation}
as the input for
(\ref{NfunctionA}),
we easily compute the corresponding GPD quintessence function.

As an example we consider the double distribution model for the quark singlet
($C=+1$)
isoscalar
($H^{(S)}_+$)
and
isovector
($H^{(V)}_+$)
combinations of nucleon GPDs.
In the forward limit
$H^{(S)}_+$
and
$H^{(V)}_+$
are reduced to the following combinations of forward quark distributions:
\begin{equation}
\begin{split}
 & H^{(S)}_+(x, \xi=0)= 
 q^{(S)}_+(x) \equiv
 u(x)+\bar{u}(x)+d(x)+\bar{d}(x)
 \,; \\&
 H^{(V)}_+(x, \xi=0)=
 q^{(V)}_+(x) \equiv
 u(x)+\bar{u}(x)-d(x)-\bar{d}(x) 
 \,.
\end{split}
\end{equation}
As the numerical input for the isoscalar and isovector forward quark distributions
$q^{(S,V)}_+(x)$
we use the LO MRST fit
($Q^2=1 \text{GeV}^2$)
\cite{Martin:2002dr}.

With the help of
(\ref{Q0final}), (\ref{Q2DD}), (\ref{Q4DD})
we perform the calculation of the
isoscalar and isovector
forward-like functions
$Q^{(S,V)}_{2\nu}(x)$
with
$\nu=0,1,2$,
which reexpress the Radyushkin double distribution Ansatz for GPDs
in the framework of the dual parametrization.
We compare the result to the general form of GPD quintessence functions
$N^{(S,V)}(x)$
for isoscalar and isovector combinations of light quark singlet ($C=+1$)
nucleon GPDs. It is also extremely instructive to compare the results
for the contributions of the several first forward-like functions
to the imaginary part of the amplitude
$\im A^{(\nu)}(\xi)$
to the exact value of
$\im A_{DD}(\xi)$.
This helps to estimate the relative importance of the non-forward
effects encoded in
$Q_{2 \nu}(x)$
with
$\nu \ge 1$
for the calculation of the amplitude.

Let us first consider the case in which the parameter $b$
of the profile function
(\ref{RadProfile})
is set to its most frequent choice
$b=1$.
For this value of
$b$
one may expect that
the small-$\xi$
asymptotic behavior of
$\im A_{DD}(\xi)$
is sufficiently well reproduced already with the help
of the first forward-like function
$Q_{0}$
since the leading small-$x$
behavior of isoscalar and
isovector forward-like distributions
$q_+^{(S,V)} \sim 1/x^{\alpha^{(S,V)}}$
is determined by the powers
$\alpha^{(S)} \sim 1$ and $0< \alpha^{(V)} <1$.

\begin{figure}[h]
 \begin{center}
  \epsfig{figure=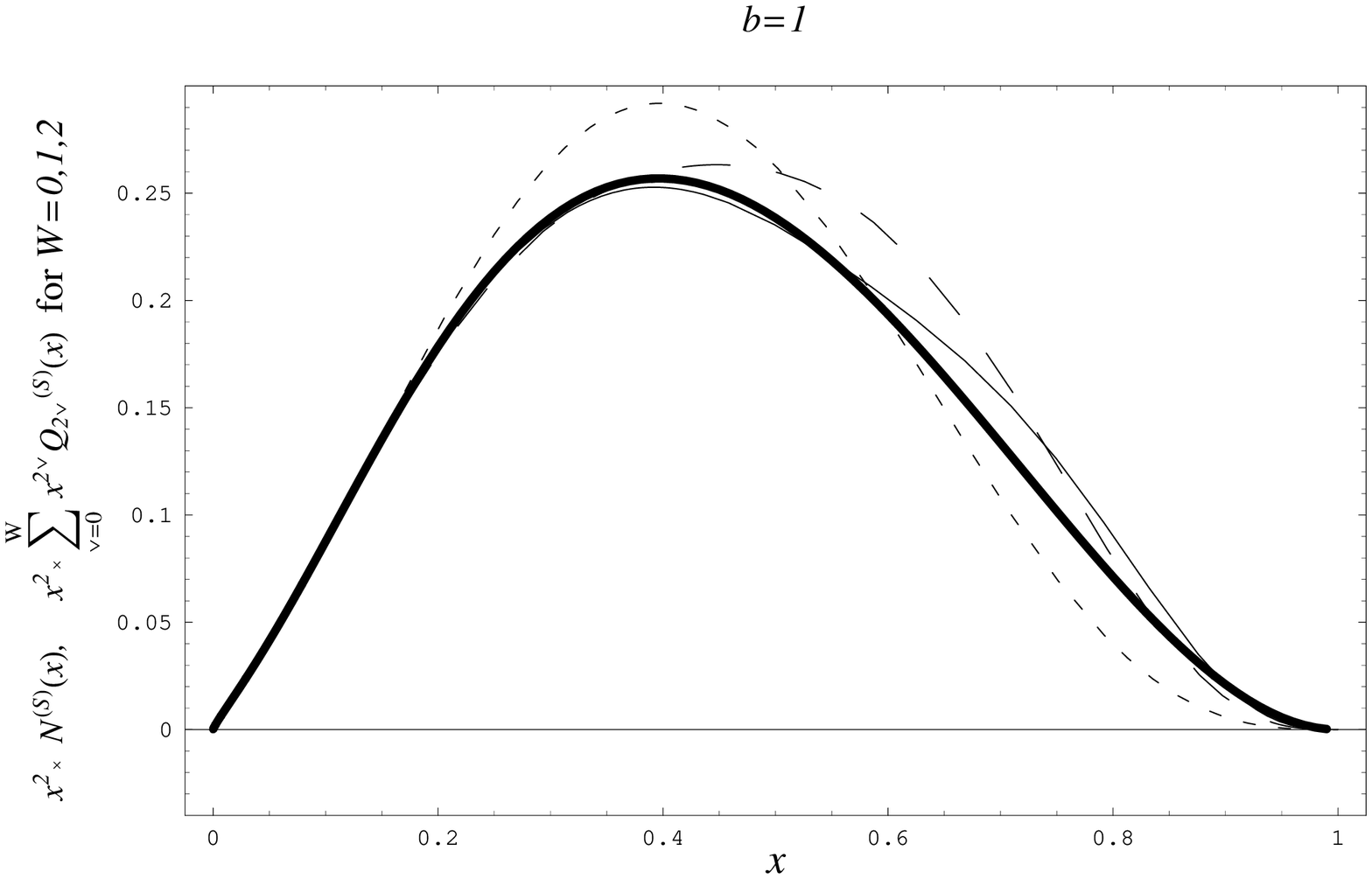, height=4.5cm}
\caption{Comparison of the consecutive contributions of the isoscalar
forward-like functions into the isoscalar GPD quintessence function
$x^2 \times Q_0^{(S)}(x)$
(dashed line with short dashes),
$x^2 \times (Q_0^{(S)}(x)+x^2 Q_2^{(S)}(x))$
(dashed line with long dashes),
$x^2\times (Q_0^{(S)}(x)+x^2 Q_2^{(S)}(x)+x^4Q_4^{(S)}(x))$
(thin solid line)
into the  isoscalar GPD quintessence function
$x^2 \times N^{(S)(x)}=x^2 \times  \sum_{\nu=0}^\infty x^{2 \nu}Q_{2 \nu}^{(S)}(x)$
(thick solid line).
}
\label{Fig1}
\end{center}
\end{figure}

In fig.~\ref{Fig1}
we show the result of an approximation of the
isoscalar
GPD quintessence function with the help of contributions of several
first forward-like functions. We compare the consecutive approximations
$\sum_{\nu=1}^W x^{2 \nu} Q_{2 \nu}(x)$
with
$W=0,1,2$
to the general result for the isoscalar (isovector) GPD quintessence
function
$N^{(S)}(x)$
for GPD in Radyushkin parametrization recovered with the help of
Abel tomography method
(\ref{NfunctionA}).
One can conclude that in this case the contribution of several first
forward-like functions really make a dominant contribution into the
GPD quintessence functions
$N^{(S)}(x)$
for small-$x$.
The same conclusion remains true for the case of the isovector
combination.

\begin{figure}
 \begin{center}
  \epsfig{figure=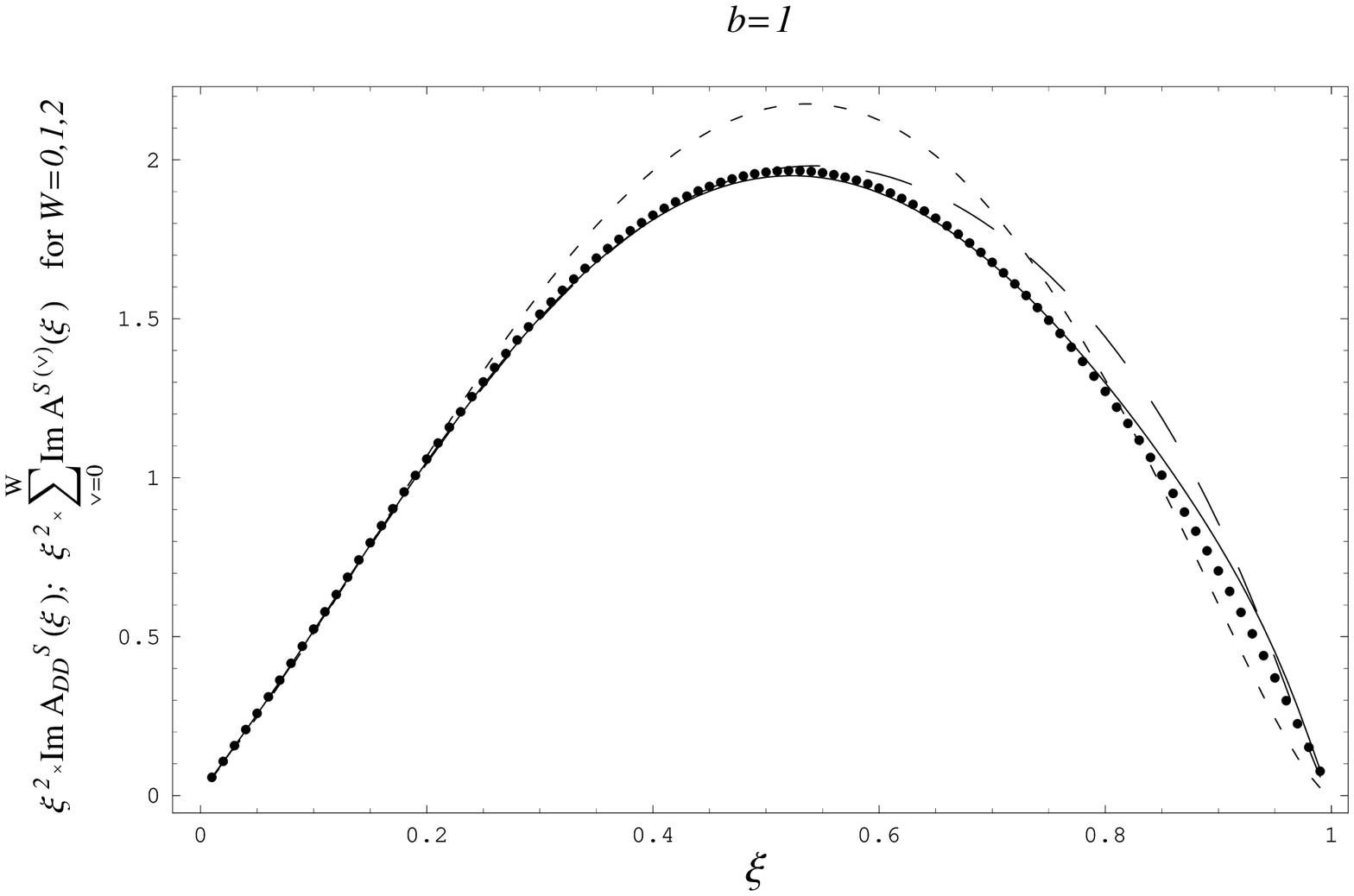, height=4.5cm}
\caption{
$\xi^2  \times \im A^S_{DD}(\xi)$
(dotted line) compared  to the result for
$\xi^2  \times \im A^{S}(\xi)$
from several first forward-like functions
$\xi^2 \times \sum_{\nu=0}^W \im A^{S \, (\nu)}(\xi)$
with
$W=0$
(short dashed line),
$W=1$
(long dashed line) and
$W=2$
(solid line).}
\label{Fig1bis}
\end{center}
\end{figure}

In fig.~\ref{Fig1bis} 
we compare the result for the imaginary part of isoscalar
DVCS amplitude
$\im A_{DD}^{S}$
to that calculated in the framework of the dual parametrization with
the help of several first forward-like functions
$Q_{2 \nu}^{(S )}$
recovered with the help of reparametrization procedure. One may check
that for
$\xi \sim 0$ the imaginary parts of isoscalar and isovector
amplitudes calculated in the framework of Radyushkin double
distribution parametrization of GPDs are with high accuracy
reproduced
in the framework of the dual parametrization of GPDs
already by the contribution of the first forward-like
function
$Q_{0}$.
However, for
$\xi \sim 1$
the contribution of several first forward-like functions
$Q_{2 \nu}$
into the imaginary part of the amplitude turns out to be
insufficient to reproduce
$\im A_{DD}^{S}$
with high accuracy, since the
behavior of
$\im A_{DD}^{S}$
for
$\xi \sim 1$
essentially differs from that of
$\im A_{DD}^{S \; (0,1,2)}$.
In order to reproduce the asymptotic behavior of
$\im A_{DD}^{S}$
in the framework of the dual parametrization one has to sum
up the whole series of contributions of
$Q_{2 \nu}$.

\begin{figure}
 \begin{center}
  \epsfig{figure=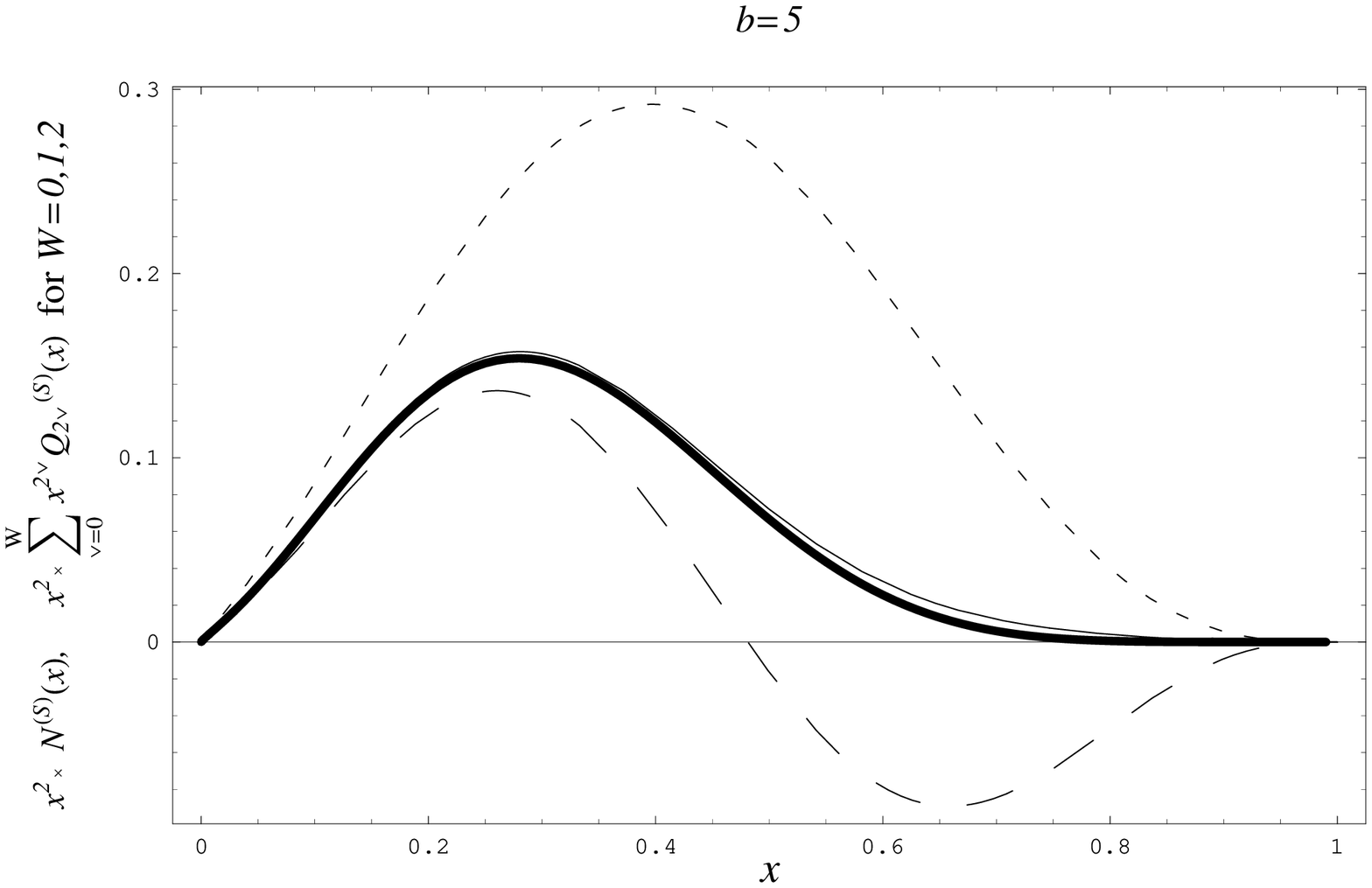, height=4.5cm}
\caption{Comparison of the consecutive contributions of the isoscalar
forward-like functions into the isoscalar GPD quintessence function
$x^2 \times Q_0^{(S)}(x)$
(dashed line with short dashes),
$x^2\times (Q_0^{(S)}(x)+x^2 Q_2^{(S)}(x))$
(dashed line with long dashes),
$x^2\times (Q_0^{(S)}(x)+x^2 Q_2^{(S)}(x)+x^4Q_4^{(S)}(x))$
(thin solid line)
into the  isoscalar GPD quintessence function
$x^2 \times N^{(S)(x)}=x^2 \times \sum_{\nu=0}^\infty x^{2 \nu}Q_{2 \nu}^{(S)}(x)$
(thick solid line).
}
\label{Fig_b5}
\end{center}
\end{figure}

\begin{figure}
 \begin{center}
  \epsfig{figure=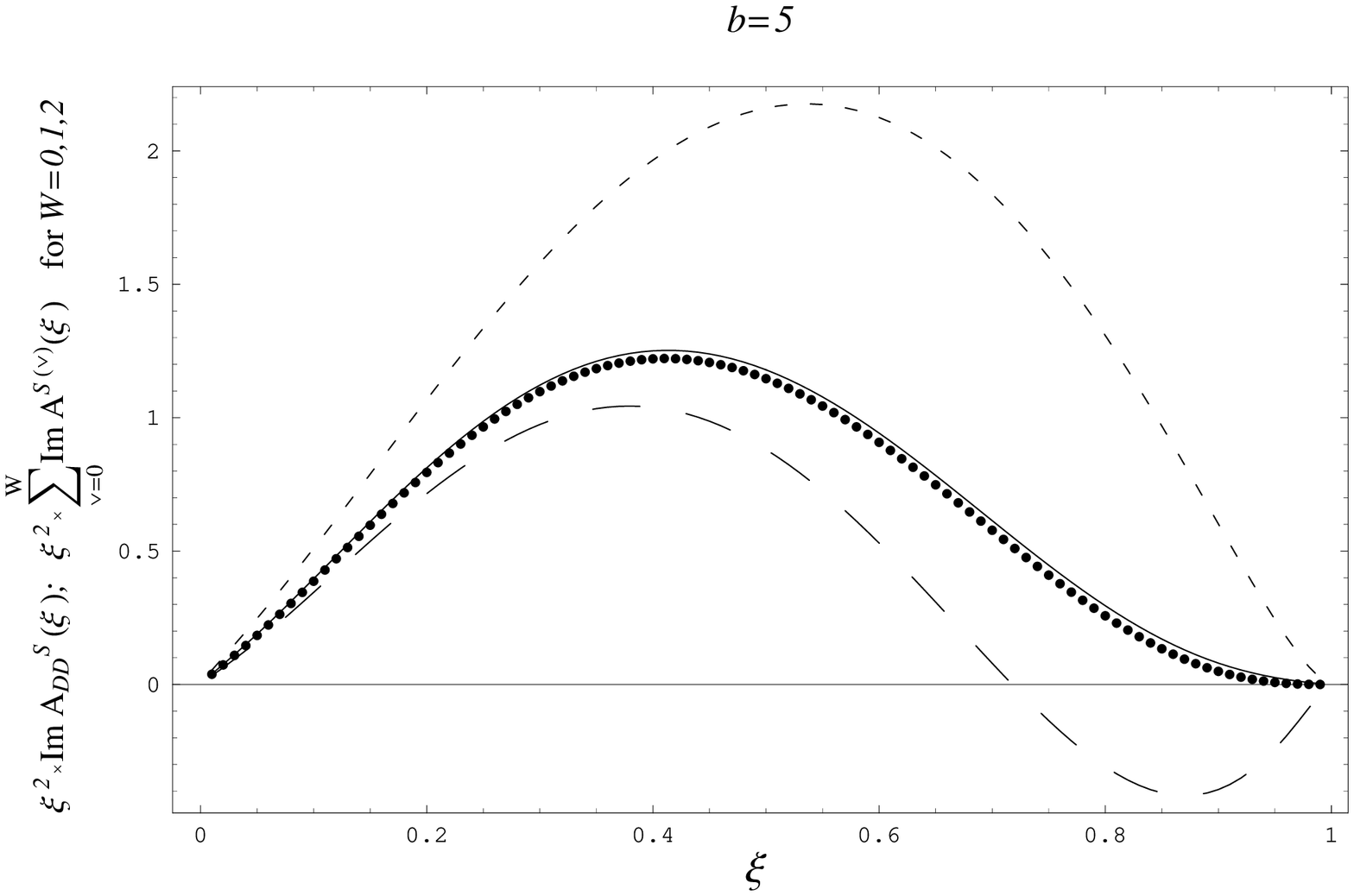, height=4.5cm}
\caption{
$\xi^2  \times \im A^S_{DD}(\xi)$
(dotted line) compared  to the result for
$\xi^2 \times \im A^{S}(\xi)$
from several first forward-like functions
$\xi^2 \times \sum_{\nu=0}^W \im A^{S \, (\nu)}(\xi)$
with
$W=0$
(short dashed line),
$W=1$
(long dashed line) and
$W=2$
(solid line).}
\label{Fig_b5_bis}
\end{center}
\end{figure}

\begin{figure}[H]
 \begin{center}
  \epsfig{figure=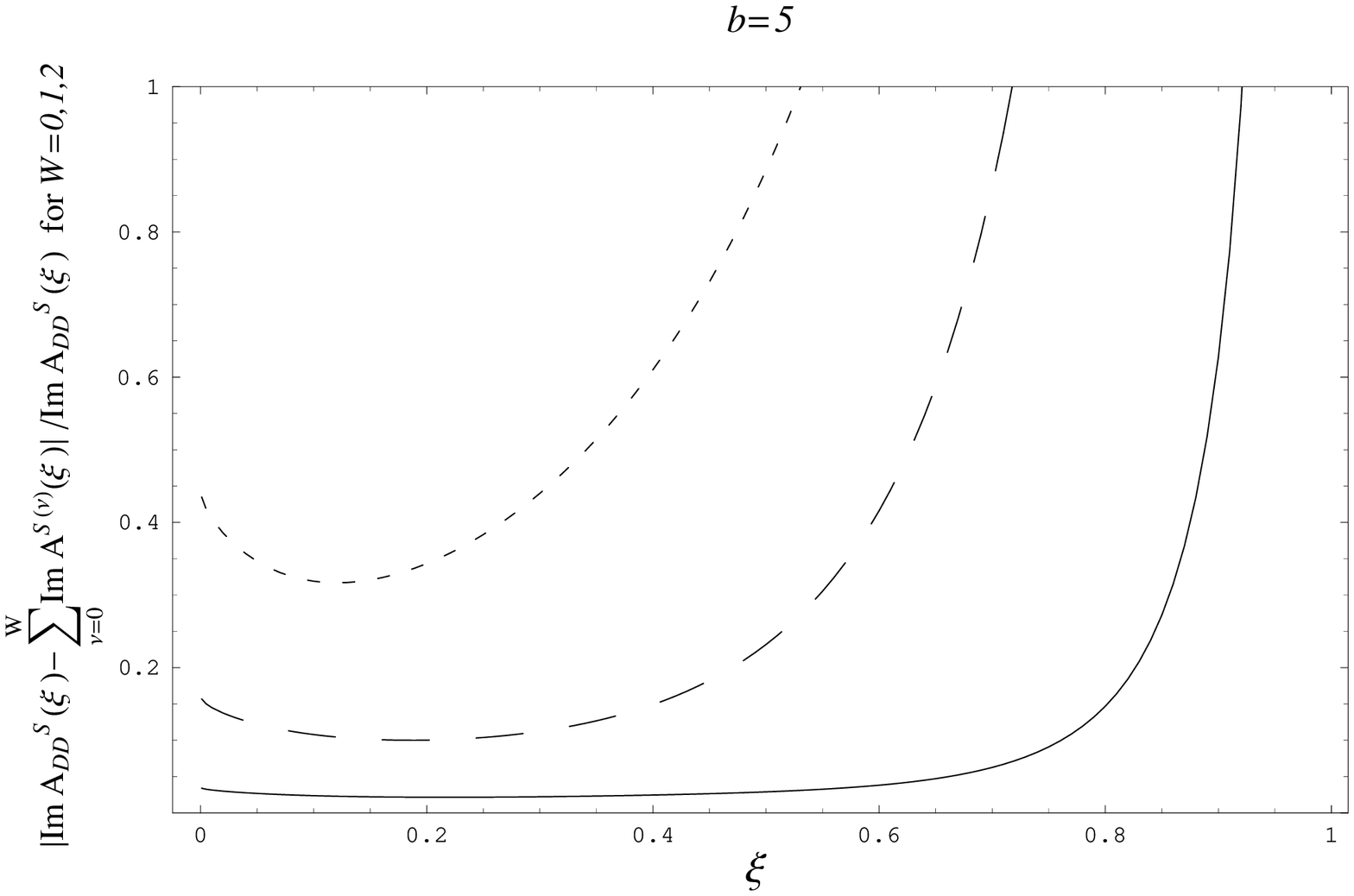 , height=4.5cm}
  \caption{Relative discrepancy between the exact result for
  $\im A^{S}_{DD}(\xi)$  and the result for  $\im A^{S}(\xi)$
  calculated in the framework of the dual parametrization
  from several first forward-like functions
  $Q_{2 \nu}$
(\ref{Q0final}),(\ref{Q2DD}), (\ref{Q4DD}):
  $\sum_{\nu=0}^W \im A^{S
  (\nu)}(\xi)$
  with $W=0$ (short dashed line),  $W=1$ (long dashed line) and $W=2$ (solid
  line).}
\label{Fig_b5_rat}
\end{center}
\end{figure}

Now we consider a different choice of the parameter $b$
of the profile function
(\ref{RadProfile}).
We set
$b=5$
and again show the results for the isoscalar case.
The GPD quintessence
function
$N^{(S)}(x)$
compared to the contribution of several first forward-like functions is
presented in fig.~\ref{Fig_b5}.
The imaginary part of the DVCS amplitude
$\im A_{DD}^{S}$
calculated in the framework
of the double distribution parametrization of GPDs
compared to the contribution of several first forward-like functions
into the imaginary part of the amplitude
is shown in fig.~\ref{Fig_b5_bis}.
Finally, in fig.~\ref{Fig_b5_rat}
we show the relative discrepancy between
$\im A_{DD}^{S}$
and the contributions of several first forward-like functions
into the imaginary part of the amplitude.
In this case taking into account only
$Q_0$
and
$Q_2$
contributions turn out to be insufficient
to reproduce the small-$\xi$
behavior of the imaginary part
$\im A_{DD}^{S}$
with high accuracy. In order
to achieve the satisfactory accuracy one has to take
into account the contributions of
$Q_0$, $Q_2$
and
$Q_4$.

Note, that the relative discrepancy depicted
in  fig.~\ref{Fig_b5_rat}
``explodes'' for
$\xi \sim 1$.
This is certainly due to the fact that the dual and double distribution
parametrizations of GPDs result in substantially different asymptotic
behavior of
$\im A(\xi)$
for
$\xi \sim 1$.
Hence, in order to reproduce accurately
$\im A_{DD}^{S}(\xi)$
in this region in the framework of the dual parametrization,
one has to sum up the whole series of the contributions of forward-like
functions that express the double distribution parametrization of GPDs
through the dual parametrization.

Finally, let us briefly consider the limiting case
$b=\infty$
that corresponds to
$H_+(x,\xi)=q_+(x)$.
In this case in order to reproduce well the asymptotical behavior
of
$\im A(\xi)$
for
$\xi \sim 0$
as well as for
$\xi \sim 1$
it is necessary to take into account the whole series of
contributions of forward-like functions
$Q_{2 \nu}$.
The corresponding GPD quintessence function receives important
contributions from
$Q_{2 \nu}$
with large
$\nu$.
This makes the straightforward reparametrization of this model for GPDs
through the dual parametrization impracticable.




\section{Conclusions}

In this paper we illustrate the application of the reparametrization
procedure, that allows to recast a particular phenomenological model
for GPDs through the dual parametrization of GPDs. We consider the popular
Radyushkin double distribution parametrization for GPDs and derive the
analytical  expressions for the corresponding forward-like functions
$Q_{2 \nu}(x)$, $\nu=1,2$.
It is interesting to note that for the most frequent choice of the
parameter
$b$
of the profile function
(\ref{RadProfile})
($b=1$)
the contribution of
$Q_{2 \nu}(x)$
with
$\nu \ge 3$
into GPD quintessence function estimated in the numerical calculation
turns out to be rather small for small-$x$ (note, that in
\cite{ForwardLikeF_KS}
the same property was revealed for the
pion GPD computed in the nonlocal chiral quark model).
Thus, for  small $x$ the corresponding GPD quintessence
function $N(x)$ is dominated by the contributions of several first
forward-like functions. We argue that this is due the fact that 
in this case the small-$\xi$
asymptotic behavior of
$\im A_{DD}(\xi)$
is well reproduced in the framework of the dual parametrization
already with help of the few first forward-like functions. For larger
values of $b$ the contribution of the forward-like functions
$Q_{2 \nu}$
with large
$\nu$
becomes more significant.

We make an important observation that the forward-like functions
$Q_{2 \nu}(x)$
with
$\nu \ge 1$
can contribute into the leading singular behavior of the
imaginary part of DVCS amplitude. This provides an opportunity for
a more flexible GPD modelling in the framework of the dual
parametrization.

We also consider a way to handle  divergencies, which may occur in
generalized form-factors
$B_{2\nu-1 \,\, 0}(t)$
due to the singularities of forward-like functions. The key to this
problem is provided by the consideration of analytic properties of
Mellin moments of GPDs in Mellin space. Once the analyticity of the Mellin
moments of GPDs is assumed, the so-called analytic regularization of
divergencies is a natural way to treat the problematical form factors.
Such approach allows to fix unambiguously the contribution of the
$D$-form factor into the real part of the DVCS amplitude in terms of
GPD quintessence function
$N(x,t)$
and the forward-like function
$Q_0(x,t)$.
On the other hand if these analyticity requirements are turned down
(e.g. by assuming the possible fixed pole contribution or explicit contact term contribution) the
value of the $D$-form factor turns out to be an independent
physical quantity that is to be fixed from the experiment.



\section*{Acknowledgements}
\mbox
We are grateful to Dieter M\"{u}ller for numerous illuminating conversations.
We are also thankful to Ya.~I.~Azimov, Alena Moiseeva, O.~Teryaev, A.~Radyushkin and M.~Vanderhaeghen
for many
valuable discussions and comments.
The work was supported by
the Sofja Kovalevskaja Programme of the Alexander von Humboldt
Foundation, Deutsche Forschungsgemeinschaft. The work of K.S. was supported by
STIBET scholarship from DAAD.

\setcounter{section}{0}
\setcounter{equation}{0}
\renewcommand{\thesection}{\Alph{section}}
\renewcommand{\theequation}{\thesection\arabic{equation}}

\section{On some properties of the integral transform
$Q_{2\nu}(x,t) \rightarrow H(x,\xi,t)$
}
\label{App_A}

In this Appendix we consider the properties of the integral transformation
relating  the singlet and nonsinglet GPDs%
\footnote{For simplicity here we consider the case of spin-$0$ target. The corresponding
formulae equally apply to the case of electric combinations of the singlet (nonsinglet)
nucleon GPDs $H_{\pm}^{(E)}(x,\xi,t)$.}
$H_{\pm}(x,\xi,t)$
to the set of forward-like functions
$Q_{2 \nu}(y,t)$.
The original derivation of this integral transformation
was presented in
Ref.~\cite{Polyakov:2002wz}
not in full details.
For example, the precise way of treating the divergensies of
the corresponding integrals has not been spelled out,
which have lead to some confusion in the literature
\cite{R.Mainz,Guzey:2005ec,Guzey:2006xi,Guzey:2008ys}.
Below we present a more detailed and accurate derivation of this
integral transformation. Our final result remains valid for the case
of small-$y$ singular behavior of the forward-like functions
$Q_{2 \nu}(y,t)$
suggested by the analysis presented in Sects.~\ref{Sec_toy}, \ref{Sec_Q2Q4_for_double}.


First we consider the case of the singlet GPD
$H_+(x,\xi)$
that is given in the framework of the dual parametrization by the series:
\be
&&
H_+(x, \xi, t)=
\nonumber \\&&
 2 \sum_{n=1 \atop \text{odd}}^\infty
\sum_{l=0 \atop \text{even}}^{n+1}
B_{nl}(t) \,
\theta
\left(
1-\frac{x^2}{\xi^2}
\right)
\left(
1-\frac{x^2}{\xi^2}
\right)
C_n^{\frac{3}{2}}
\left( \frac{x}{\xi} \right)
P_l
\left( \frac{1}{\xi} \right)\,.
\nonumber
\\
\label{FSGPD_singlet}
\ee
The formal series (\ref{FSGPD_singlet}) corresponds to the
analytic continuation of the corresponding well convergent
expansion for the generalized distribution amplitudes entering the
description of $\gamma^* \gamma \rightarrow h \bar{h}$ hard process
to the cross channel (see discussion in \cite{Polyakov:1998ze}).
The crucial point is that in the region
$0 \le \xi\le 1$
the representation
(\ref{FSGPD_singlet})
for GPD is to be understood as the one which
formally satisfies the polynomially condition
($N=1,\,3,\,5,\,...$):
\be
&&
\int_{0}^1 dx \, x^N H_+(x,\xi,t)= \nonumber \\&&
  h_0^{(N)}(t)+h_2^{(N)}(t) \xi^2+...+h_{N+1}^{(N)}(t) \xi^{N+1}\,,
\ee
where the coefficients
$h_{2 \nu}^{(N)}(t)$
(with $\nu \le \frac{N+1}{2}$)
at powers of
$\xi$
are given by
the finite sums
\begin{equation}
\begin{split}
  &
  h_{2 \nu}^{(N)}(t)=\sum_{n=1 \atop \text{odd}}^N
  \sum_{l=0 \atop \text{even}}^{n+1}
  B_{nl}(t) \,
  {(-1) }^{\frac{2 \nu +l-N-1 }{2}} \\&
  \times
    \frac{\Gamma (1 - \frac{2 \nu - l - N}{2})}{\Gamma (\frac{1}{2} + \frac{2 \nu + l - N}{2})\,\Gamma (2 - 2 \nu + N)\,}
    \\&
       \times
    \frac{\left(  n+1 \right) \,\left(  n+2 \right) \,\Gamma ( N+1)\,
    }{2^{2 \nu}\,
    \Gamma (1 + \frac{N-n}{2})\,\Gamma (\frac{5}{2} + \frac{N + n}{2})}\,.
\end{split}
\end{equation}

The general idea of the method
\cite{Polyakov:2002wz}
employed for the summation of the
formal partial wave expansion for GPD
consists in presenting GPD as a result of convolution of a certain
kernel with the set of  forward-like functions
$Q_{2 \nu}(y,t)$ $(\nu=0, \,1,\,...)$
whose Mellin moments generate the generalized form factors
$B_{nl}(t)$:
\be
B_{n \, n+1-2\nu}(t)= \int_0^1 dy y^n
Q_{2\nu}(y,t)\,. 
\label{Bnl}
\ee
The explicit construction of this convolution kernel allows one to derive
the rigorous expressions for GPDs in the framework of the dual parametrization.


We start with the definition of the discontinuity of the particular function
$f(z)$:
\begin{equation}
{\rm disc}_{z=x}\,f(z)= \frac{1}{2 \pi i}
\left(
f(x- i0)-f(x+i0)
\right)
\end{equation}
The basic relation derived in
\cite{Polyakov:2002wz}
reads
\be
&&
{\rm disc}_{z=x} \, \frac{1}{y}
\left(1 +y \frac{\partial}{\partial y}
\right)
\int_{-1}^1 ds \, z_s^{-N} \nonumber \\&&=
\theta \left(1 -\frac{x^2}{\xi^2} \right)
\xi^{-N} y^{N-1} C_{N-1}^{3/2} \left( \frac{x}{\xi} \right)\,,
\label{disc_block}
\ee
where
\begin{equation}
z_s= 2 \frac{z-s \xi}{(1-s^2)y}\,,
\end{equation}
with
$0 \le y \le 1$.
%

Let us now consider the following function:
\be
&&
F^{(2 \nu)}(z,y) \nonumber \\&&=
\frac{1}{y}
\left(
1+y \frac{\partial}{\partial y}
\right)
\int_{-1}^1 ds \, \xi^{2 \nu} z_s^{1-2 \nu}
\frac{1}{\sqrt{z_s^2-2 z_s +\xi^2}}\,. \nonumber \\
\ee
The discontinuity of this function
is a suitable building block for the convolution kernel whose
convolution with the set of forward-like functions
$Q_{2 \nu}$
allows to reproduce the formal series
(\ref{FSGPD_singlet}).
Let us explicitly compute the discontinuity of this function
employing the well known property of the generating function
of the system of Legendre polynomials:
\be
&&
\frac{1}{\sqrt{\xi^2 -2 z_s+z_s^2}}=
\frac{1}{z_s}
\frac{1}{\sqrt{\left(\frac{\xi}{z_s} \right)^2- 2 \frac{1}{z_s}+1 }}
\nonumber \\&&
=
\sum_{l=0}^\infty P_l \left( \frac{1}{\xi} \right)
\xi^l \left( \frac{1}{z_s} \right)^{l+1}
\label{expansion_met1}
\ee
Employing
(\ref{disc_block})
we obtain
\be
&& \nonumber
{\rm disc}_{z=x} F^{(2 \nu)}(z,y)  \\&& =
{\rm disc}_{z=x} \frac{1}{y} \left(1+ y \frac{\partial}{\partial y} \right)
\int_{-1}^1 ds \,\xi^{2 \nu} \sum_{l=0}^\infty \xi^l  z_s^{-2 \nu-l} P_l \left( \frac{1}{\xi} \right)
\nonumber
\\&&
=\left(
1-\frac{x^2}{\xi^2}
\right)
\theta
\left(
1-\frac{x^2}{\xi^2}
\right)
\nonumber \\&&
\times
{\sum_{l=0}^{\infty}}^* C_{2 \nu+l-1}^{\frac{3}{2}}
\left(
\frac{x}{\xi}
\right)
P_l \left( \frac{1}{\xi} \right) y^{2 \nu+l-1}\,.
\nonumber \\
\label{Kernel_Fk}
\ee
The asterisk in the sum in the last line of (\ref{Kernel_Fk})
denotes that for $\nu=0$ the term with
$\nu=l=0$
is actually absent.
Let us consider the following integral convolution:
\be
\sum_{\nu=0 }^\infty
\int_{0}^1 dy \,
&&
\left\{
{\rm disc}_{z=x} F^{(2 \nu)}(z,y)
\right.
\nonumber \\&&
\left.
-
{\rm disc}_{z=-x} F^{(2 \nu)}(z,y)
\right\}
Q_{2 \nu}(y,t)\,.
\label{Convolution_Master}
\ee
Now using
(\ref{Kernel_Fk})
together with the expressions for the generalized form factors
(\ref{Bnl})
we obtain
\be
&&
\nonumber
\sum_{\nu=0}^\infty
\int_{0}^1 dy \,
\left\{
{\rm disc}_{z=x} F^{(2 \nu)}(z,y)
\right.
\\&&
\left.
-
{\rm disc}_{z=-x} F^{(2 \nu)}(z,y)
\right\}
Q_{2 \nu}(y,t) \nonumber \\&&
\nonumber
=\theta \left( 1- \frac{x^2}{\xi^2} \right)
\left( 1- \frac{x^2}{\xi^2} \right)
\\&&
\nonumber
\times
2 
\sum_{\nu=0 }^\infty
{\sum_{l=0 \atop \text{even}}^\infty}^*
C^{\frac{3}{2}}_{2 \nu+l-1} \left( \frac{x}{\xi} \right)
P_l \left( \frac{1}{\xi} \right) B_{2 \nu+l-1 \;\; l}(t) \\&&
=
\theta \left( 1- \frac{x^2}{\xi^2} \right)
\left( 1- \frac{x^2}{\xi^2} \right) \nonumber \\&&
\times
2
\sum_{n=1 \atop \text{odd}}^\infty
{\sum_{l=0 \atop \text{even}}^{n+1}}
C^{\frac{3}{2}}_{n} \left( \frac{x}{\xi} \right)
P_l \left( \frac{1}{\xi} \right) B_{n \;l}(t)\,,
\label{Kernel_Fk_details}
\ee
where in the last line we have interchanged the order of summation
introducing the new summation index
$n \equiv 2 \nu+l-1$.

The trick is that we can compute the discontinuity of the function
$F^{(2 \nu)}(z,y)$
with the help of an alternative method. Namely,
instead of using the formal expansion
(\ref{expansion_met1})
we can consider the contribution into discontinuity
$F^{(2 \nu)}(z,y)$
stemming from the cut
$1-\sqrt{1-\xi^2} < z_s<1+\sqrt{1-\xi^2}$ and from
the poles at
$z_s=0$
for
$\nu \ge 1$.

Let start with specifying the contribution of the cut.
According to the standard definition of the discontinuity
of a real analytic function ($f^*(z)= f(z^*)$)  with a branch
cut along the real axis:
\be
&&
{\rm disc}_{z=x} f(z)= \frac{1}{2 \pi i}
\left(
f(x- i0)-f^*(x-i0)
\right)
\nonumber \\&&
= \frac{1}{\pi}  \; {\rm Im} \, f(x-i0)
\ee
Thus, the discontinuity of
$F^{(2 \nu)}(z,y)$
can be computed as
\be
&&
{\rm disc}_{z_s^2-2 z_s+\xi^2=x_s^2-2 x_s+\xi^2} F^{(2 \nu)}(z,y)
\nonumber \\&&
= \frac{\xi^{2 \nu}}{\pi} \,
{\rm Im}
\left\{
\frac{1}{y} \left(
1+y \frac{\partial}{\partial y}
\right)
\int_{-1}^1 ds \, \frac{x_s^{1-2 \nu}}{\sqrt{x_s^2-2 x_s+\xi^2-i \epsilon}}
\right\} \nonumber \\&&
+
\;
\left\{
{ \text{pole} \atop \text{contrib}\,.}
\right\}\,. \nonumber \\
\label{disc_stage0}
\ee
Special attention is to be paid to the choice of the physical
branch of the square root in
(\ref{disc_stage0}).
According to our convention
\be
{\rm Im} \frac{1}{\sqrt{x-i \epsilon}}=
{\rm Im} \frac{1}{\sqrt{|x|}}\, e^{i \frac{\pi}{2}}=
 \frac{1}{\sqrt{ |x| }} \;
\theta (-x)
\label{brach_choice}
\ee
%
Now let us discuss in details the contribution to the discontinuity
of
$F^{(2 \nu)}(z,y)$
resulting from poles at
$z_s=0$.
Employing the expansion
\be
\frac{1}{\sqrt{\xi^2 -2 z_s+z_s^2}}= \sum_{l=0}^\infty P_l \left( \frac{1}{\xi} \right)
z_s^l \left( \frac{1}{\xi} \right)^{l+1}
\ee
we can present
$F^{(2 \nu)}(z,y)$
as follows:
\be
&&
F^{(2 \nu)}(z,y)=
\frac{1}{y} \left(1+ y \frac{\partial}{\partial y} \right) \nonumber \\&&
\int_{-1}^1 ds \,\sum_{l=0}^\infty \xi^{2 \nu-l-1}
 (z_s+i \epsilon)^{1-2 \nu+l} P_l \left( \frac{1}{\xi}
\right) \,.
\ee
Obviously, for a given
$\nu$
only a finite number of pole terms arise with
$l<2 \nu-1$.
Note, that our choice of
$i \epsilon$
prescription is here matched with our
convention
(\ref{brach_choice})
for the physical branch of the square root in
(\ref{disc_stage0}).
Thus, we conclude that
(\ref{disc_stage0})
may be rewritten as
\be
&&
\int_0^1 dy \, Q_{2 \nu}(y,t) \,{\rm disc}_{z=x} F^{(2 \nu)}(z,y) \nonumber \\&&
= \frac{\xi^{2 \nu}}{\pi}
\int_0^1 dy \, Q_{2 \nu}(y,t) \frac{1}{y}
\left(
1+y \frac{\partial}{\partial y}
\right) \nonumber \\&&
\int_{-1}^1 ds \frac{x_s^{1-2 \nu}}{\sqrt{2x_s-x_s^2-\xi^2}}
\,
\theta(2x_s-x_s^2-\xi^2)
 \nonumber \\&&
-\,
\theta\left(1- \frac{x^2}{\xi^2}\right)
\left(1- \frac{x^2}{\xi^2}\right) \nonumber \\&&
\times
\sum_{l=0}^{2 \nu-2}
C_{2 \nu-l-2}^{\frac{3}{2}}
\left( \frac{x}{\xi} \right)
P_l
\left( \frac{1}{\xi} \right)
\int_0^1 dy y^{2 \nu-l-2} Q_{2 \nu}(y,t) \nonumber \\
 \label{disc_stage1}
\ee
As it was noted in
\cite{Tomography},
the sign in front of the last term in
(\ref{disc_stage1})
actually differs from that stated in the original paper
\cite{Polyakov:2002wz}.

In order to proceed further one needs to perform the analysis of the solutions of the
algebraic equation
\be
2 x_s -x_s^2-\xi^2=0\,,
\label{eq_xs}
\ee
where
$x_s= 2 \frac{x- \xi s}{(1-s^2)y}$.
The four roots of the equation
(\ref{eq_xs})
$s_i$, ($i=1,...\,4$)
are given by the following expressions:
\begin{equation}
\begin{split}
& s_1=\frac{1}{y}
\left(
\mu  - \sqrt{ \left( 1 - x\,y \right) \,\left( 1 + {\mu }^2 \right)-(1 - y^2 )
}
\right); \\&
s_2=\frac{1}{y}
\left(
\mu  + \sqrt{ \left( 1 - x\,y \right) \,\left( 1 + {\mu }^2 \right)-(1 - y^2 )
}
\right);
\\&
s_3=\frac{1}{y}
\left(
\lambda  - \sqrt{ \left( 1 - x\,y \right) \,\left( 1 + {\lambda }^2 \right)-(1 - y^2 )
}
\right);
\\&
s_4=\frac{1}{y}
\left(
\lambda  + \sqrt{ \left( 1 - x\,y \right) \,\left( 1 + {\lambda }^2 \right)-(1 - y^2 )
}
\right).
\end{split}
\label{roots_s14}
\end{equation}
Here we have employed the notations
\begin{equation*}
\mu= \frac{1-\sqrt{1-\xi^2}}{\xi}\,, \ \ \ \ \lambda=\frac{1}{\mu}\;.
\end{equation*}
The solutions of the equation
$s_1=s_2$
are given by
$y=y_0$ and $y=\frac{1}{y_1}$
while
$y=y_1$ and $y=\frac{1}{y_0}$ are the solutions of the equation
$s_3=s_4$.
The expressions for $y_{0,1}$ read as
\begin{equation}
y_0=
\frac{x\,\left( 1 + {\mu }^2 \right) }{2} +
{\sqrt{  \frac{x^2\,{\left( 1 + {\mu }^2 \right) }^2}{4}-{\mu }^2 }};
\label{Y0}
\end{equation}
\begin{equation}
y_1=
\frac{x\,\left( 1 + {\lambda }^2 \right) }{2} +
{\sqrt{  \frac{x^2\,{\left( 1 + {\lambda }^2 \right) }^2}{4}-{\lambda }^2 }}.
\label{Y1}
\end{equation}

This allows to rewrite
(\ref{disc_stage1})
as
\be
&&
\int_0^1 dy \, Q_{2 \nu}(y,t) \,{\rm disc}_{z=x} F^{(2 \nu)}(z,y) \nonumber \\&&
=\xi^{2 \nu}
\left\{
\theta(x>\xi)
\int_{y_0}^1 \frac{dy}{y}
\left[
\left(1- y \frac{\partial}{\partial y}\right)Q_{2 \nu}(y,t)
\right]
\right.
\nonumber \\&&
\left.
\times
\frac{1}{\pi}
\int_{s_1}^{s_2} ds \frac{x_s^{1-2 \nu}}{\sqrt{2x_s-x_s^2-\xi^2}}
\right.
\nonumber \\&&
\left.
+ \theta(-\xi<x<\xi)
\int_{0}^1 dy \, Q_{2 \nu}(y,t)
\frac{1}{y}
\left(
1+y \frac{\partial}{\partial y}
\right)
\right.
\nonumber
\\&&
\left.
\left[
\frac{1}{\pi}
\int_{s_1}^{s_3} ds \frac{x_s^{1-2 \nu}}{\sqrt{2x_s-x_s^2-\xi^2}}
\right.
\right.
\nonumber \\&&
\left.
\left.
-\frac{1}{\xi^{2 \nu}}
\left(1- \frac{x^2}{\xi^2} \right)
\sum_{l=0}^{2 \nu-2}
C_{2 \nu-l-2}^{\frac{3}{2}}
\left(\frac{x}{\xi} \right)
P_l
\left(\frac{1}{\xi} \right)
\frac{y^{2 \nu-l-1}}{2 \nu-l}
\right]
\right\}
\nonumber \\
\label{convolution_stepII}
\ee

A special attention is to be payed to the convergency of the
overall integral in
$y$
in the second term of
(\ref{convolution_stepII}).
For this we need to consider the small $y$
asymptotic behavior of the the elliptic integral
\be
\frac{1}{\pi}
\int_{s_1}^{s_3}
ds
\frac{x_s^{1-2 \nu}}{\sqrt{2x_s-x_s^2-\xi^2}}\,.
\label{int_s1s3}
\ee
For
$x \in (-\xi; \, \xi)$
the following asymptotic behavior
of the integral
(\ref{int_s1s3})
for
$y \sim 0$
can be established:
\be
&&
\nonumber
\frac{1}{\pi}
\int_{s_1}^{s_3}
ds
\frac{x_s^{1-2 \nu}}{\sqrt{2x_s-x_s^2-\xi^2}}
\\ &&
=
\frac{1}{\xi^{2 \nu}}
\left(1- \frac{x^2}{\xi^2} \right)
\sum_{l=0}^{2 \nu-2}
C_{2 \nu-l-2}^{\frac{3}{2}}
\left(\frac{x}{\xi} \right)
P_l
\left(\frac{1}{\xi} \right)
\frac{y^{2 \nu-l-1}}{2 \nu-l}
\nonumber
\\ &&
+
\frac{1}{\xi^{2 \nu}}
\left(1- \frac{x^2}{\xi^2} \right)
\nonumber \\&&
\times
\sum_{l=-2}^{\min{(-1,2 \nu-2)}}
C_{2 \nu-l-2}^{\frac{3}{2}}
\left(\frac{x}{\xi} \right)
P_l
\left(\frac{1}{\xi} \right)
\frac{y^{2 \nu-l-1}}{2 \nu-l}
+
O(y^{2 \nu+2})\,,
\nonumber \\&&
\label{stage_x_disc}
\ee
where
$P_{-n}(\chi) \equiv P_{n-1}(\chi)$.
The first term in
(\ref{stage_x_disc})
is exactly cancelled by the pole contribution in
(\ref{convolution_stepII}).
The term with
$l=-1$
in the second sum in
(\ref{stage_x_disc})
that is $O(y^{2 \nu})$
contributes solely to the $D$-term.
Finally,
the term with
$l=-2$
in the second sum in
(\ref{stage_x_disc})
that is
$O(y^{2 \nu+1})$
is even in
$x$
and therefore does not survive in the singlet combination.

We also have to specify our assumptions concerning
the small
$y$
singular behavior of forward-like functions
$Q_{2 \nu}(y,t)$.
The singular behavior of the singlet forward-like function
$Q_0(y,t)$
is determined by that of the singlet combination of
the corresponding forward quark distributions:
\be
Q_0(y) \sim \frac{1}{y^{\alpha}} \ \ \ \text{with} \ \ \ 1<\alpha<2\,.
\label{asymp_Q0}
\ee
We argue that the reasonable singular behavior of
$Q_{2 \nu}(y)$
with $\nu>0$
for small
$y$
is given by
\be
Q_{2 \nu}(y) \sim \frac{1}{y^{2 \nu+\alpha}} \ \ \ \text{with} \ \ \ 1<\alpha<2\,.
\label{asymp_Qk}
\ee

The next step is to add and subtract the combination
\be
&&
\frac{\pi}{\xi^{2 \nu}}
\left(1- \frac{x^2}{\xi^2} \right)
\sum_{l=-2}^{\min{(-1,2 \nu-2)}}
C_{2 \nu-l-2}^{\frac{3}{2}}
\left(\frac{x}{\xi} \right)
P_l
\left(\frac{1}{\xi} \right)
\frac{y^{2 \nu-l-1}}{2 \nu-l}
\nonumber \\ &&
\ee
from the elliptic integral in the second term of
(\ref{convolution_stepII}).
Now integrating by parts according to
\be
&&
\left.  \int_0^1 dy \, Q(y) \frac{\partial}{\partial y} \phi(y) \right.
\nonumber \\ &&
\left.
=-
\int_0^1 dy  \left[ \frac{\partial}{\partial y} Q(y) \right] \phi(y)+
Q(y) \phi(y) \right|_{0}^1
\label{by_parts}
\ee
and employing the asymptotic conditions
(\ref{asymp_Q0}), (\ref{asymp_Qk})
we can rewrite the general expression
(\ref{Convolution_Master})
for the singlet GPD
in terms of the forward-like functions
as follows:
\begin{equation}
\begin{split}
& H_+(x,\xi,t)=
\sum_{\nu=0}^\infty
\int_{-1}^1 dy \,
\left\{
{\rm disc}_{z=x} F^{(2 \nu)}(z,y)
\right.
\\&
-
\left.
{\rm disc}_{z=-x} F^{(2 \nu)}(z,y)
\right\}
Q_{2 \nu}(y,t) \\&
=\sum_{\nu=0}^\infty
\xi^{2 \nu}
\left[
H^{( \nu)}_+(x,\xi,t)-H^{( \nu)}_+(-x,\xi,t)
\right]\,
\\&
+2 
\sum_{\nu=1}^\infty
\theta \left( 1- \frac{x^2}{\xi^2}\right)
\left( 1- \frac{x^2}{\xi^2}\right)
C_{2 \nu-1}^{\frac{3}{2}}
\left(
\frac{x}{\xi}
\right)
B_{2 \nu -1 \; 0}(t)\,,
\end{split}
\label{H_dual_through_Qk}
\end{equation}
where the functions
$H^{( \nu)}_+(x, \xi, t)$
defined for
$-\xi \le x \le 1$
are given by the following integral transformations:
\be
 && H^{(\nu)}_+(x,\xi,t)=
\theta(x>\xi)
\frac{1}{\pi}
\int_{y_0}^1 \frac{dy}{y}
\left[
\left(
1-y \frac{\partial}{\partial y}
\right)
Q_{2 \nu}(y,t)
\right]
\nonumber \\&&
\times \int_{s_1}^{s_2} ds\, \frac{x_s^{1-2 \nu}}{\sqrt{2 x_s-x_s^2-\xi^2}}
\nonumber
\\&&
+ \theta(-\xi<x<\xi)
\frac{1}{\pi}
\int_{0}^1 \frac{dy}{y}
\left[
\left(
1-y \frac{\partial}{\partial y}
\right)
Q_{2 \nu}(y,t)
\right]
\nonumber \\&&  \times
\left\{
\int_{s_1}^{s_3} ds \frac{x_s^{1-2 \nu}}{\sqrt{2 x_s-x_s^2-\xi^2}}-
 \frac{\pi}{\xi^{2 \nu}}
\left(
1- \frac{x^2}{\xi^2}
\right)
\right.
\nonumber \\&&
\left.
\times
\sum_{l=-2}^{2 \nu -2}
C_{2 \nu -l-2}^{\frac{3}{2}}
\left(
\frac{x}{\xi}
\right)
P_l
\left(
\frac{1}{\xi}
\right)
\frac{y^{2 \nu-l-1}}{2 \nu-l}
\right\}
\,, \nonumber \\
\label{Hk_main}
\ee
with
$P_{-n}(\chi) \equiv P_{n-1}(\chi)$.
Note that the integral over $y$ in the second term
of (\ref{Hk_main}) is  well convergent under the assumptions
(\ref{asymp_Q0}), (\ref{asymp_Qk})
(see that the terms in braces in the third line of
(\ref{Hk_main})
behave as
$y^{2 \nu +2}$ for
$y \sim 0$).
We also stress that in this case
the terms outside the integral stemming from
(\ref{by_parts})
advocated in \cite{R.Mainz,Guzey:2006xi,Guzey:2008ys}
happily vanish.
The only part of
(\ref{H_dual_through_Qk})
that still may suffer from divergensies is
the second sum in (\ref{H_dual_through_Qk})
containing the generalized form factors
$B_{2 \nu -1 \; 0}(t)$:
\be
B_{2 \nu -1 \; 0}(t)= \int_0^1 \frac{dy}{y} \, y^{2 \nu } Q_{2 \nu}(y,t)\,.
\label{problematicFF}
\ee
An important observation is that the second sum in
(\ref{H_dual_through_Qk})
is a pure $D$-term contribution.
As it is explained in
sect.~\ref{Sec_toy},
in order to ensure the convergency of integrals
in (\ref{problematicFF}),
in this case it turns
out necessary to introduce a regularization
(see discussion in sect.~\ref{Sec_toy}).

Our final expression
(\ref{H_dual_through_Qk})
for the
singlet GPD through the set of forward-like functions
differs from that presented in literature
\cite{Polyakov:2002wz,Tomography}.
The reason for this is that the  results
\cite{Polyakov:2002wz,Tomography}
were derived under
the assumption
that
\be
\lim_{y \rightarrow 0} y^2 Q_0(y,t)=0; \ \ \ \
\lim_{y \rightarrow 0} y^{2 \nu} Q_{2 \nu}(y,t)=0, \ \ \ \nu>0\,. \nonumber \\
\label{asymp_cond}
\ee
The first condition in
(\ref{asymp_cond})
is certainly respected in our case,  while according to the analysis
presented in
sects.~\ref{Sec_toy}, \ref{Sec_Q2Q4_for_double}
the second one seems to be too restrictive.
However, it is straightforward to check  that under the assumptions
(\ref{asymp_cond})
the result
(\ref{H_dual_through_Qk}), (\ref{Hk_main})
for
$H_+(x,\xi,t)$
is reduced to that presented in
\cite{Tomography}.

We also present the summary of formulae for the case of the
nonsinglet
$(C=-1)$
GPD $H_-(x,\xi,t) \equiv H^q(x,\xi,t)+H^q(-x,\xi,t)$.
In the limit
$\xi \rightarrow 0$
$H_-(x,\xi,t)$
is reduced to $q_-(x,t) \equiv q(x,t)- \bar{q}(x,t)$.
The following partial wave expansion for
$H_-(x,\xi,t)$
can be written in the framework of the dual parametrization:
\be
&&
H_-(x, \xi,t)=
\nonumber \\&&
2 \sum_{n=0 \atop \text{even}}^{\infty}
\sum_{l=1 \atop \text{odd}}^{n+1}
B_{nl}(t)
\theta \left(1 -\frac{x^2}{\xi^2} \right)
\left(1 -\frac{x^2}{\xi^2} \right)
C_n^{\frac{3}{2}} \left( \frac{x}{\xi} \right)
P_l \left( \frac{1}{\xi} \right)\,. \nonumber \\
\label{PWave_H_NS}
\ee
In order to sum the formal series
(\ref{PWave_H_NS})
we introduce the set of the nonsinglet forward-like functions
$Q_{2 \nu}(y,t)$
whose Mellin moments give the generalized form factors
$B_{n \, l}(t)$
analogously to
(\ref{Bnl}).
The reasonable singular behavior of the nonsinglet
forward-like functions
$Q_{2 \nu}(y)$
with
$\nu>0$
for small
$y$
is given by
\be
Q_{2 \nu}(y) \sim \frac{1}{y^{2 \nu+\alpha_-}} \ \ \ \text{with} \ \ \ 0<\alpha_-<1\,.
\label{asymp_Qkns}
\ee
Then. the following integral transform relating the nonsinglet GPD
$H_-(x,\xi,t)$
to the set of the nonsinglet forward-like functions
$Q_{2 \nu}(y)$
can be established:
\be
&& H_-(x,\xi,t)=\sum_{\nu=0}^\infty
\int_{-1}^1 dy \,
\left\{
{\rm disc}_{z=x} F^{(2 \nu)}(z,y)
\right.
\nonumber \\&&
\left.
+
{\rm disc}_{z=-x} F^{(2 \nu)}(z,y)
\right\}
Q_{2 \nu}(y,t)
\nonumber \\
&&
=\sum_{\nu=0 }^\infty
\xi^{2 \nu}
\left[
H^{(\nu)}_-(x,\xi,t)+H^{(\nu)}_-(-x,\xi,t)
\right]\,,
\ee
where
$H^{\nu}_-(x,\xi,t)$
defined for $-\xi \le x \le 1$
is given by
\be
 && H^{( \nu)}_-(x,\xi,t)=
\theta(x>\xi)
\frac{1}{\pi}
\int_{y_0}^1 \frac{dy}{y}
\left[
\left(
1-y \frac{\partial}{\partial y}
\right)
Q_{2 \nu}(y,t)
\right] \nonumber \\ && \times
\int_{s_1}^{s_2} ds\, \frac{x_s^{1-2 \nu}}{\sqrt{2 x_s-x_s^2-\xi^2}}
\nonumber
\\&&
+ \theta(-\xi<x<\xi)
\frac{1}{\pi}
\int_{0}^1 \frac{dy}{y}
\left[
\left(
1-y \frac{\partial}{\partial y}
\right)
Q_{2 \nu}(y,t)
\right]
\nonumber \\&&  \times
\left\{
\int_{s_1}^{s_3} ds \frac{x_s^{1-2 \nu}}{\sqrt{2 x_s-x_s^2-\xi^2}}
\right.
\nonumber \\&&
\left.
-
 \frac{\pi}{\xi^{2 \nu}}
\left(
1- \frac{x^2}{\xi^2}
\right)
\sum_{l=-1}^{2 \nu -2}
C_{2 \nu -l-2}^{\frac{3}{2}}
\left(
\frac{x}{\xi}
\right)
P_l
\left(
\frac{1}{\xi}
\right)
\frac{y^{2 \nu-l-1}}{2 \nu-l}
\right\}
\,, \nonumber \\
\label{Hk_main_NS}
\ee
with
$P_{-n}(\chi) \equiv P_{n-1}(\chi)$.

\section{Explicit expression for $Q_4(x)$ corresponding to the double distribution
parametrization of GPD $H_+(x,\xi)$}
\label{Expl_Q4_DD}

In this Appendix we present the explicit expression for
the forward-like function
$Q_4(x)$
calculated from matching of small
$\xi$
expansion of GPD $H_+$
in the framework of the dual parametrization
and
(\ref{Small_xi_Rad_Exp}):
\begin{equation}
Q_4(x)=f_{Q_4}^{(b)}(x)+ \int_x^1 dy \, K_{Q_4}^{(b)}(x,y) \, q_+(y),
\end{equation}
where
\be
&&
f_{Q_4}^{(b)}(x)=
\left(
\frac{-525 + 2660\,x - 5310\,x^2 + 5700\,x^3 - 2909\,x^4}{2\,( 5 + 2\,b )}
\right.
\nonumber \\&&
+\left.
\frac{1365 - 4788\,x + 6478\,x^2 - 4692\,x^3 + 1893\,x^4}{2\,( 3 + 2\,b )}
\right.
\nonumber
\\&&
\left.
+
( 49 - 17\,x^2 )
( 1 - x^2) \frac{}{}
\right) \frac{}{}
\frac{1}{8 x^4} \, q_+(x)  \nonumber \\&&
+\left(
\frac{ -35 + 185\,x - 625\,x^2 + 859\,x^3 }
{ 5 + 2\,b  }
\right.
\nonumber \\&&
\left.
+
\frac{ 91 - 337\,x + 625\,x^2 - 603\,x^3 }
{ 3 + 2\,b }
+ \left( 7 - 11\,x^2 \right) \left( 1 + x \right)
\right)
\nonumber \\&&
\times
\frac{(1-x)}{4x^3} \,
q_+'(x) \nonumber \\&&
+\left(
-\frac{  5 - 70\,x + 209\,x^2    }
  { 5 + 2\,b }+
\frac{ 17 - 94\,x + 165\,x^2  }{ 3 + 2\,b }+
(1+x)^2
\right)
\nonumber \\&&
\times
\frac{(1-x)^2}{2x^2}\,
q_+''(x) \nonumber \\&&
+\left( \frac{   -5 + 37\,x  }{  5 + 2\,b }+
\frac{3\,\left( 3 - 11\,x \right) }{3 + 2\,b }
\right) \frac{(1-x)^3}{2x}
q^{(3)}_+(x)
\nonumber \\&&
+
\left(
- \frac{1 } {5 + 2\,b}
+\frac{1 } {3 + 2\,b}
\right)
(1-x)^4 \,
q^{(4)}_+(x)
\ee
and the convolution kernel is given by
\begin{equation}
\begin{split}
& K_{Q_4}^{(b)}(x,y) \\& =-\frac{1}{16\,y^6} \left( 315\,x + 140\,y - 270\,x\,y^2 - 84\,y^3 + 27\,x\,y^4 \right)
 \\&+
\frac{105}{32\,\left( 5 + 2\,b \right) \,y^6}
\left( 45\,x + 4\,y - 180\,x\,y - 16\,y^2
\right. \\&
\left.
+ 270\,x\,y^2 + 24\,y^3- 180\,x\,y^3 - 16\,y^4 +
       45\,x\,y^4 + 4\,y^5 \right)\\&-
  \frac{3}{32\,\left( 3 + 2\,b \right)
       \,y^6}
\left( 3675\,x + 700\,y - 10500\,x\,y \right. \\& \left. - 1680\,y^2   + 11010\,x\,y^2 + 1344\,y^3 -
       5220\,x\,y^3 \right. \\&
       \left. - 448\,y^4 + 1035\,x\,y^4 + 84\,y^5 \right).
\end{split}
\end{equation}

%


%
%

\end{document}